\documentclass[preprint, 3p]{elsarticle}
\journal{MSSP}
\usepackage{hyperref}
\hypersetup{
 pdftitle    = {pdftitle},
 pdfsubject  = {pdfsubject},
 pdfauthor   = {pdfauthor},
 pdfkeywords = {pdfkexwords},
 pdfborder   = 0 0 0,
 plainpages  = false,
 bookmarksnumbered = true
}

\usepackage{amssymb,amsmath,amsfonts,amsbsy}
\usepackage{color}
\usepackage{url}
\usepackage[shortlabels]{enumitem}
\usepackage{xspace}
\usepackage{siunitx}
\usepackage{mathtools}

\usepackage{bm}

\usepackage{graphicx}        % standard LaTeX graphics tool

\usepackage[utf8]{inputenc}

\usepackage{subcaption}
\usepackage{caption}
\usepackage[beginLComment=,endLComment=]{algpseudocodex}

\usepackage{pgfplots}
\pgfplotsset{compat=newest}
\pgfplotsset{plot coordinates/math parser=false}
\newlength\figW
\newlength\figH
\newcommand{\nc}{\newcommand}
\newcommand{\rnc}{\renewcommand}
\nc{\bs}{\boldsymbol}
\nc{\mm}{\bs}
\nc{\RM}[1]{\MakeUppercase{\romannumeral #1}}
\nc{\mrm}[1]{\mathrm{#1}}
\nc{\ie}{i.\,e.\xspace}
\nc{\eg}{e.\,g.\xspace}
\nc{\etal}{et al.\xspace}
\nc{\mbf}{\mathbf}
\nc{\forceVec}{\bs{f}}
\nc{\fabstand}{\,}
\nc{\fp}{\fabstand.}
\nc{\fk}{\fabstand,}
\nc{\fref}[1]{{Fig.~\ref{f:#1}}}
\nc{\sref}[1]{{Section~\ref{s:#1}}}
\nc{\ssref}[1]{{Section~\ref{s:#1}}}
\nc{\sssref}[1]{{Section~\ref{s:#1}}}
\nc{\aref}[1]{{\ref{a:#1}}}
\nc{\eref}[1]{{Eq.~\ref{e:#1}}}
\nc{\erefs}[1]{{Eqs.~\ref{e:#1}}}
\nc{\erefo}[1]{{\ref{e:#1}}}
\nc{\ereft}[2]{{Eqs.~\ref{e:#1}~and~\ref{e:#2}}}
\nc{\erefm}[2]{{Eqs.~\ref{e:#1}-\ref{e:#2}}}
\nc{\tref}[1]{{Table~\ref{t:#1}}}
\nc{\hs}[1]{\hspace*{#1}}
\nc{\vs}[1]{\vspace*{#1}}
\nc{\x}[1]{\mbox{#1}}
\nc{\prog}[1]{{\sf{#1}}\xspace}
\nc{\name}[1]{\textsc{#1}\xspace}
\rnc{\phi}{\varphi}
\rnc{\theta}{\vartheta}
\rnc{\rho}{\varrho}
\rnc{\epsilon}{\varepsilon}
\nc{\dd}{{\mathrm{d}}}
\nc{\ii}{{\mathrm{i}}}
\nc{\ee}{{\mathrm{e}}}
\nc{\lk}{(\hs{1ex})}
\nc{\g}[1]{\x{$#1$}}
\nc{\mex}{m_{\mrm{ex}}}
\nc{\dex}{d_{\mrm{ex}}}
\nc{\kex}{k_{\mrm{ex}}}
\nc{\omex}{\omega_{\mrm{ex}}}
\nc{\Dex}{D_{\mrm{ex}}}
\nc{\phex}{\phi_{\mrm{ex}}}
\nc{\muex}{\mu_{\mrm{ex}}}
\nc{\eex}{{\mm e}_{\mrm{ex}}}
\nc{\qex}{q_{\mrm{ex}}}
\nc{\dqex}{\dot{q}_{\mrm{ex}}}
\nc{\ddqex}{\ddot{q}_{\mrm{ex}}}
\nc{\kp}{k_{\mrm p}}
\nc{\ki}{k_{\mrm i}}
\nc{\zi}{z_{\mrm i}}
\nc{\dzi}{\dot{z}_{\mrm i}}
\nc{\kpnd}{\bar{k}_{\mrm p}}
\nc{\kind}{\bar{k}_{\mrm i}}
\nc{\err}{\epsilon}
\nc{\thref}{\frac\pi2}%{\vartheta_{\mrm{ref}}}
\nc{\thf}{\vartheta_f}
\nc{\thfest}{\hat{\vartheta}_f}
\nc{\dthfest}{\dot{\hat{\vartheta}}_f}
\rnc{\th}{\vartheta}
\nc{\dth}{\dot{\vartheta}}
\nc{\thest}{\hat{\vartheta}}
\nc{\dthest}{\dot{\hat{\vartheta}}}
\nc{\thdel}{\vartheta_\Delta}
\nc{\thdelest}{\hat{\vartheta}_\Delta}
\nc{\Arg}[1]{\operatorname{Arg}\left\{ #1 \right\}}
\nc{\Uc}{U}
\nc{\delpl}{\delta_{\mrm{p}}}
\nc{\dellin}{\delta_{\mrm{s}}}
\nc{\phexlin}{\phi_{\mrm{ex,lin}}}
\nc{\omlin}{\omega_{\mrm{lin}}}
\nc{\muexlin}{\mu_{\mrm{ex,lin}}}
\nc{\omLP}{\omega_{\mrm{LP}}}
\nc{\lamR}{\lambda_\mathrm{R}}
\nc{\lamI}{\lambda_\mathrm{I}}
\nc{\delplnd}{\bar{\delpl}}
\nc{\omnd}{\bar{\omega}}
\nc{\lamRnd}{{\lambda}_\mathrm{R}}
\nc{\lamInd}{{\lambda}_\mathrm{I}}
\nc{\errtol}{\err_{\mrm{tol}}}
\rnc{\matrix}[2]{
\left[\!\!\begin{array}{#1}#2\end{array}\!\!\right]}
\rnc{\vector}[1]{\matrix{c}{#1}}
\nc{\ea}[1]{
\begin{eqnarray}
#1 \end{eqnarray}}
\nc{\e}[2]{\begin{equation} #1 \label {e:#2} \end{equation}}
\nc{\real}[1]{\Re\left\{ #1\right\}}
\nc{\imag}[1]{\Im\left\{ #1\right\}}
\definecolor{red}{rgb}{1.0,0.0,0.0}
\definecolor{green}{rgb}{0.0,1.0,0.0}
\definecolor{blue}{rgb}{0.0,0.0,1.0}
\definecolor{black}{rgb}{0.0,0.0,0.0}
\definecolor{cyan}{rgb}{0.0,1.0,1.0}
\definecolor{magenta}{rgb}{1.0,0.0,1.0}
\definecolor{matlab1}{rgb}{0, 0.4470, 0.7410}
\definecolor{matlab2}{rgb}{0.8500, 0.3250, 0.0980}
\definecolor{matlab3}{rgb}{0.9290, 0.6940, 0.1250}
\definecolor{matlab4}{rgb}{0.4940, 0.1840, 0.5560}
\definecolor{greyg}{rgb}{0.7,0.7,0.7}
\nc{\MOD}[1]{#1}
\newcommand*{\ccline}[2]{{#2}~(\textcolor{#1}{\rule[0.5ex]{0.5cm}{0.55pt}})}
\newcommand*{\ccarea}[2]{{#2}~(\textcolor{#1}{\rule[0ex]{0.5cm}{1ex}})}

\makeindex             % used for the subject index

\begin{document}
%=============================================================================
% TITLE and ABSTRACT
%=============================================================================
\begin{frontmatter}
%\title{(Towards systematic) design of phase-resonance testing}
\title{Robust and fast backbone tracking via phase-locked loops}
%\title{How to rapidly and robustly track backbones via phase-locked loops?}
% KEYWORDS FOR TITLE
% - phase resonance testing | backbone tracking
% - (systematic) design
% - phase-locked loop
% - adaptive filtering based on LMS algorithm
\author{
Patrick Hippold$^1$,
Maren Scheel$^1$,
Ludovic Renson$^2$
Malte Krack$^1$
}
%\address{email addresses}
\address{$^1$ University of Stuttgart, GERMANY}
\address{$^2$ Imperial College London, UK}
%\address{email addresses}

\begin{abstract}
Phase-locked loops are commonly used for shaker-based backbone tracking of nonlinear structures.
The state of the art is to tune the control parameters by trial and error.
In the present work, an approach is proposed to make backbone tracking much more robust and faster.
A simple PI controller is proposed, and closed-form expressions for the gains are provided that lead to an optimal settling of the phase transient.
The required input parameters are obtained from a conventional shaker-based linear modal test, and an open-loop sine test at a single frequency and level.
For phase detection, an adaptive filter based on the LMS algorithm is used, which is shown to be superior to the synchronous demodulation commonly used.
Once the phase has locked, one can directly take the next step along the backbone, eliminating the hold times.
The latter are currently used for recording the steady state, and to estimate Fourier coefficients in the post-process, which becomes unnecessary since the adaptive filter yields a highly accurate estimation at runtime.
The excellent performance of the proposed approach is demonstrated for a doubly clamped beam undergoing bending-stretching coupling leading to a 20 percent shift of the lowest modal frequency.
Even for fixed control parameters, designed for the linear regime, only about 100 periods are needed per backbone point, also in the nonlinear regime.
This is much faster than what has been reported in the literature so far.
%In fact, the nonlinear backbone test becomes faster than the linear modal test, shifting the established paradigm.
%, and makes backbone tracking suitable for slowly time-variable systems (e.g. due to thermal transients).
\end{abstract}

\begin{keyword}
phase resonance \sep PLL \sep NNM \sep force appropriation
\end{keyword}
\end{frontmatter}

\noindent \fbox{
%\color{blue}
\begin{minipage}[c]{\textwidth-13.47pt}
\section*{List of symbols}
\begin{minipage}[t]{0.55\textwidth}
\subsection*{Time, frequency, phase}
\begin{tabular}{ll}
     % time
     $t$        & time \\
     $\Omega$   & frequency \\
     $\tau$     & phase
\end{tabular}
\subsection*{Excitation and response}
\begin{tabular}{ll}
     $u$        & voltage \\
     $U$        & voltage amplitude\\
     $f$        & applied force \\
     $F$        & Fourier coefficient of $f$ \\
     $q_{\mrm{ex}}$ &  drive point displacement \\
     $Q$        & Fourier coefficient of $q_{\mrm{ex}}$
\end{tabular}
\subsection*{Exciter physical model}
\begin{tabular}{ll}
     % electrical
     $i$        & current \\
     $R$        & resistance \\
     $G$        & electromotive force constant \\
     % exciter mechanical
     $k_{\mrm{ex}}$        & mechanical exciter stiffness \\
     $d_{\mrm{ex}}$        & mechanical exciter damping \\
     $m_{\mrm{ex}}$        & moving exciter mass
\end{tabular}
\subsection*{Structure physical model}
\begin{tabular}{ll}
     $\mm q$    & vector of generalized coordinates \\
     $\bs{M}$   & mass matrix \\
     $\bs{g}$   & restoring and damping forces \\
     $\bs{e}_{\mrm{ex}}$   & applied force direction
\end{tabular}
\subsection*{Plant modal model}
\begin{tabular}{ll}
     % plant modal
     $\omega$   & modal frequency \\
     $D$            & modal damping ratio \\
     $\bs{\phi}$    & mass-normalized modal deflection shape \\
     $a$            & modal amplitude \\
     $\theta$       & phase shift \\
     $\delpl$       & decay rate of plant \\
     $\dellin$      & decay rate of structure under test\\
     $\mu_{\mrm{ex}}$ & mass ratio exciter/structural mode
\end{tabular}
\end{minipage}
\begin{minipage}[t]{0.43\textwidth}
\subsection*{Adaptive filter}
\begin{tabular}{ll}
     % adaptive filter
     $H$            & harmonic truncation order \\
     $\omLP$    & low-pass filter cutoff frequency
\end{tabular}
\subsection*{Controller}
\begin{tabular}{ll}
     % controller
     $\Omega_\mrm{ini}$ & initial frequency of phase-locked loop \\
     $\err$     & phase shift error \\
     $\errtol$  & tolerance specified for $\err$ \\
     $\kp$          & gain of proportional controller part \\
     $\ki$          & gain of integral controller part \\
     $I_\err$       & integral state of phase controller
\end{tabular}
\subsection*{Linearized state space model}
\begin{tabular}{ll}
     % linearized state-space model
     $\bs{z}$        & state vector \\
     $\bs{A}_0$     & matrix of linearized dynamical system \\
     $\Upsilon$     & constant of plant \\
     $\lambda$      & eigenvalue
\end{tabular}
\subsection*{Constants}
\begin{tabular}{ll}
     % general constants
     $\ii$          & imaginary unit \\
     $\ee$          & Euler number
\end{tabular}
\subsection*{Annotations}
\begin{tabular}{ll}
    $\phantom{\Delta} \dot{\square}$     & derivative with respect to $t$ \\
    $\phantom{\Delta} \bar{\square}$     & non-dimensionalized with $\omLP$ \\
    $\phantom{\Delta} \square^\prime$    & derivative with respect to $\bar{t}$ \\
    $\phantom{\Delta} \tilde{\square}$   & including exciter effects\\
    $\phantom{\Delta} \hat{\square}$     & estimation via adaptive filter \\
    $\phantom{\Delta} \square^{(h)}$     & $h$-th harmonic \\
    $\phantom{\Delta} \square_\mrm{ex}$  & at drive point/of the exciter \\
    $\phantom{\Delta} \square_\mrm{lin}$ & linear(-ized) \\
    $\phantom{\Delta} \square_\mrm{R}$   & real part \\
    $\phantom{\Delta} \square_\mrm{I}$   & imaginary part \\
    $\Delta \square$    & deviation from fixed point
\end{tabular}
\end{minipage}
\end{minipage}
}

%%%%%%%%%%%%%%%%%%%%%%%%%%%%%%%%%%%%%%%%%%%%%%%%%%%%%
\section{Introduction\label{s:intro}}
% systematic literature review ...
% ( TITLE-ABS-KEY ( ( "phase resonance" AND vibr* ) OR nnm OR "nonlinear mod*" OR "nonlinear normal mod*" ) AND TITLE-ABS-KEY ( "phase-locked loop" ) AND TITLE-ABS-KEY ( test* OR experiment* ) ) AND ( LIMIT-TO ( DOCTYPE , "ar" ) OR LIMIT-TO ( DOCTYPE , "bk" ) )
% ... did not yield additional relevant sources
%%%%%%%%%%%%%%%%%%%%%%%%%%%%%%%%%%%%%%%%%%%%%%%%
The goal of the present work is to achieve a fast and robust tracking of the phase-resonant backbone curve of nonlinear structures, by a systematic design of the phase-locked loop feedback controller.
To motivate the problem setting, an overview is given in the following on the main use cases of phase resonance testing.
Then, different means of achieving phase resonance are discussed, including the use of phase-locked loops.
Subsequently, the state of the art of the design of phase-locked loops for vibration testing is reviewed.
Finally, the outline of the present work is described.

\subsection*{What is phase resonance testing and what is it used for?}
% LINEAR
Phase resonance testing is mainly used for the identification of modal properties, both in the linear and in the nonlinear case.
In the linear case, the approach is commonly referred to as tuned-sinusoidal or force appropriation (vibration) testing, and it is commonly used to separate modes with closely-spaced frequencies.
Here, besides finding the appropriate frequency, the key challenge is to determine an appropriate force pattern (multi-shaker testing) to bring the response into phase resonance at each exciter location, and thus to isolate the individual modes.
\\
% NONLINEAR
Phase resonance testing is also well-known for its ability to investigate (weakly) nonlinear behavior \cite{ewin1995}.
Atkins \etal \cite{Atkins.2000} showed that a multi-harmonic forcing is theoretically needed in the nonlinear case.
Peeters \etal \cite{Peeters.2011,Peeters.2011b} revisited \cite{Atkins.2000} and concluded that single-point mono-harmonic forcing often permits a good isolation of a nonlinear mode, which is an important finding from a practical perspective.
A requirement is that damping must be light and the modal frequency must be well-separated (there are no strong modal interactions).
It is useful to note that there are different definitions of nonlinear modes.
The goal of \cite{Peeters.2011,Peeters.2011b} was to identify a Nonlinear Normal Mode, which is defined a family of periodic motions of the underlying conservative autonomous (unforced-undamped) system.
A definition of nonlinear modes that explicitly accounts for (possibly nonlinear) dissipation is the Extended Periodic Motion Concept \cite{Krack.2015a}.
To isolate such a mode, the phase-resonant backbone (forced-damped configuration) should be tracked as shown in \cite{Scheel.2018,Muller.2022}.
For finite damping and closely-spaced modes, the backbone of the unforced-undamped system may deviate considerably from the phase-resonant backbone \cite{Renson.2018}.
% , and also from the curve connecting the amplitude maxima of the frequency response as function of the excitation level
Besides the identification of amplitude-dependent modal properties, backbone tracking can be useful to uncover isolated frequency response branches, as shown numerically \cite{Kuether.2015} and experimentally \cite{Woiwode.2024}.

\subsection*{How to achieve phase resonance? What are phase-locked loops?}
% PHASE RESONANCE USING VELOCTIY FEEDBACK
To achieve phase resonance, manual tuning of the excitation frequency was used in \cite{Peeters.2011b}.
As an alternative, this can be achieved using feedback control.
A simple example is a velocity feedback loop.
Here, the velocity is measured and fed back (via a gain) to the structure under test\footnote{
In some cases, the sign of the velocity is taken before it is fed back \cite{SOKOLOV.2001}.
However, the sign function introduces higher harmonics and is sensitive to noise.
This may distort the resonant dynamics severely compared to a mono-harmonic excitation, especially if the structure does not behave like a single-degree-of-freedom system \cite{SebastianMojrzisch.2012}.
}.
%In practice, this may work fine for systems that behave as single-degree-of-freedom systems \cite{SebastianMojrzisch.2012}.
%Otherwise, it is not straight-forward to select the mode which should be driven into resonance.
In the case of velocity feedback, it is not trivial to select the mode to be driven into resonance.
To enable a mode selection under a single exciter location, one can acquire the velocity at multiple locations, and include a modal filter into the feedback loop \cite{Scheel.2022}. % NOTE: direct velocity feedback (without sign function)
%They achieved this by feeding back the sign of the time-shifted velocity.
%In practice, this may work fine for systems that behave as single-degree-of-freedom systems \cite{SebastianMojrzisch.2012}.
%For multi-degree-of-freedom systems, two problems arise.
%First, the sign function introduces higher harmonics and is sensitive to noise.
%This may distort the dynamics severely compared to a mono-harmonic excitation.
%Second, it is not straight-forward to select the mode which should be driven into primary resonance.
%To achieve such a selection for a single exciter location, one can acquire the response at multiple locations, and use a modal filter \cite{Scheel.2022x}. % NOTE: direct velocity feedback (without sign function)
%The phase-locked loop does not have those difficulties, but it generally requires a controller which needs to be properly tuned.
\\
% CBC-BASED vs. PLL-BASED PHASE CONTROL
Two alternatives to velocity feedback are Control-Based Continuation and the use of phase-locked loops.
Compared to velocity feedback, both techniques permit an easier mode selection, and do not require that the response is measured at multiple locations.
An important benefit of Control-Based Continuation is its robustness; its most important downside is its inherently iterative character \cite{Renson.2016b,Abeloos.2022}.
Phase-locked loops, on the other hand, do not require any iterations, so they have the potential for faster backbone tracking, and this is why they are the focus of the present work.
%%%
Remarkably, phase resonance testing using a phase-locked loop was already applied in the 1970s to identify natural frequencies of biological tissue \cite{Axelsson.1976}. % 10.1111/j.1399-3054.1976.tb03919.x
Another common application is fatigue testing of MEMS resonators, where the natural frequency varies with time as the crack growths, and the phase-locked loop permits to maintain resonant operation \cite{Connally.1993}. %[Connally, J. A., and Brown, S. B., 1993, Micromechanical Fatigue Testing, Exp. Mech., 33~2!, pp. 81–90.]
%%%
Phase-locked loops are well-known in electrical and control engineering, and they recently gained popularity for nonlinear vibration testing.
A phase-locked loop adjusts the frequency input to the structure under test until a specified phase lag of the response is reached. % (fundamental harmonic of the)
Besides phase-resonance testing, phase-locked loops are useful also for frequency response testing, where phase control can increase robustness and even obtain coexisting frequency responses (including those that are unstable under open-loop conditions) \cite{SOKOLOV.2001}.
% MORE ON PHASE CONTROL
%Near resonance, the response is highly sensitive to the frequency.
%When the structure is driven in the nonlinear regime, turning points may arise leading to coexisting vibration states at the same frequency.
%On the other hand, the vibration states (near a primary resonance) typically have a unique phase lag between (the fundamental harmonics of) response and excitation.
%To increase robustness and even recover coexisting frequency responses, phase control can be useful, as demonstrated by Sokolov and Babitsky \cite{SOKOLOV.2001}.
%%
%They achieved this by feeding back the sign of the time-shifted velocity.
%In practice, this may work fine for systems that behave as single-degree-of-freedom systems \cite{SebastianMojrzisch.2012}.
%For multi-degree-of-freedom systems, two problems arise.
%First, the sign function introduces higher harmonics and is sensitive to noise.
%This may distort the dynamics severely compared to a mono-harmonic excitation.
%Second, it is not straight-forward to select the mode which should be driven into primary resonance.
%To achieve such a selection for a single exciter location, one can acquire the response at multiple locations, and use a modal filter \cite{Scheel.2022x}. % NOTE: direct velocity feedback (without sign function)
%The phase-locked loop does not have those difficulties, but it generally requires a controller which needs to be properly tuned.

\subsection*{What is the state of the art design of phase-locked loops for vibration testing?}
% PHASE DETECTORS
Synchronous demodulation (also known as homodyne detection) is by far the most popular means of estimating (or detecting) the phase lag during phase-controlled vibration testing \cite{Denis.2018,Scheel.2022}. % Scheel.2020, Schwarz.2020, Abeloos.2022, Muller.2022, Muller.2023 <-- seem relevant here, but we should not have too many self-citations
The input signal (force or response) is multiplied by the sine / cosine of the instantaneous phase.
The latter is obtained as the integral of the frequency output of the controller.
By applying a low-pass filter, one obtains an estimate of the sine / cosine Fourier coefficient of the signal.
In earlier works, the sign of the signal was used instead of the signal itself within the phase detector \cite{Peter.2017,Scheel.2018}.
This might permit simpler digital implementations, and inherently makes the output level independent of the input signal level \cite{Twiefel.2008}.
However, it usually performs poorly for phase controlled nonlinear vibration testing, since it is highly sensitive to noise and higher harmonics in the signal.
Denis et al. \cite{Denis.2018} introduced separate phase detectors for force and response, and use the two-argument arctangent to compute the phase lag, which also ensures an output level independent of the input.
% fourth order Butterworth filter in experiment instead of first-order filter
%- extension of \cite{Fan.2007} to phase distinct from resonance (still SDOF Duffing); stability conditions more complex and dependent, among others, on forcing level
In the present work, an adaptive filter based on the LMS algorithm \cite{Widrow.1975} is used for phase detection, and more generally, for Fourier decomposition.
This is commonly used to remove periodic disturbances from a given signal (adaptive notch filter).
In the present context, the filtering property of the LMS algorithm is not used, but the fact that it gives a steady-flow estimate of the discrete Fourier transform \cite{Widrow.1987} is particularly suitable for phase and amplitude control tasks within nonlinear vibration testing.
%Actually, the filter property is not really used, but the fact that the algorithm gives . % [Fundamental Relations Between the LMS Algorithm and the DFT]
% NOTE: factor 2*mu used in \cite{Widrow.1975} (Eq. A.15) in contrast to \cite{Abeloos.2021}!
%This seems ideal for phase and amplitude control tasks within nonlinear vibration testing.
It was introduced already by Abeloos \etal \cite{Abeloos.2021} to this context, and found superior to synchronous demodulation in \cite{Abeloos.Thesis}.
\\
% DESIGN OF CONTROLLER [phase-resonance|...] testing
In the context of vibration testing based on phase-locked loops, simple controllers are commonly used (integral, proportional-integral, or proportional–integral–derivative controllers).
A stability condition was established for the case of a pure integral controller in \cite{Sun.2002,Fan.2007,Denis.2018}, and for a proportional–integral–derivative controller in \cite{Abeloos.Thesis}.
For this purpose, the plant was modeled as a linear single-degree-of-freedom system \cite{Sun.2002}, with an additional cubic spring \cite{Fan.2007,Denis.2018}, or with a more generic nonlinear term \cite{Abeloos.Thesis}.
In all those studies, an averaging formalism was used, and the asymptotic behavior around the locked state was considered.
For the case of a pure integral controller, a maximum integral gain was established, beyond which the closed loop diverges \cite{Sun.2002,Fan.2007,Denis.2018}.
This limit depends mainly on the plant damping and the cutoff frequency of the loop filter (within the phase detector), and was found to be independent of the cubic spring \cite{Fan.2007,Denis.2018}.
It should be emphasized that the asymptotic stability is only a necessary criterion for a good controller design, and it does not say anything about the robustness or the duration of the settling time.
It was observed that an additional proportional gain reduces the settling time down to a certain limit value \cite{Peter.2017}.
In all studies on vibration testing known today, the phase-locked loop parameters were set in a strictly heuristic way.

\subsection*{What alternatives are available for backbone tracking?}
Techniques have also been proposed to identify backbone curves, or more specifically, amplitude-dependent modal frequencies and damping ratios, from the free decay (ring down response), see \eg \cite{Feldman.2011,Dion.2013b,Londono.2015}.
If the initial excitation is applied via an impact hammer, the response is generally comprised of multiple modes, and an accurate estimation of individual backbones can only be expected in the rather weakly nonlinear regime, see \eg \cite{Deaner.2015}.
If the initial excitation is provided by a shaker, instantaneously switching off the excitation (\eg by detaching structure and stinger) without distorting the dynamics of interest is very difficult or even impossible (\eg in the case of base excitation).
\\
Instead of directly tracking the phase-resonant backbone curve, one can test the frequency-response surface (or manifold) in the space spanned by frequency, response amplitude, excitation amplitude and response-excitation phase lag.
The most common ways to do this are to step the frequency while keeping the response amplitude at a target value via feedback control (Response Controlled stepped-sine Testing), see \eg \cite{Karaagacl.2021}, or to keep the frequency constant while stepping the target value for the response amplitude, to generate so-called S-curves, see \eg \cite{Abeloos.2022}.
From the gathered data points, the backbone curve can be obtained by interpolation.
Those techniques collect data that is unnecessary when focusing on backbone curves alone.
%In contrast to the direct tracking of the backbone curve, those techniques require one to generate (many) more than one point per amplitude, which results in spur
More data commonly means longer test duration, and since the goal of the present work is to obtain the backbone quickly, the focus is placed on direct backbone tracking.

\subsection*{Outline of the present work}
%=============================================================================
% PURPOSE OF PRESENT WORK
%=============================================================================
%- all parameters, in particular, filter properties (order, cutoff frequency), P and I gains of controller, systematically designed
%- based on shaker-based linear modal test + open-loop sine test at single frequency and level; no (further) knowledge, in particular no mathematical model, needed
%motivation:
%    - less time necessary for preparation (heuristic tuning)
%    - shorter duration of the actual test; shorter excitation of the specimen at resonance/high amplitudes (risk of damage / life consumption); handling slowly time-variant properties (e.g. influence of temperature)
%\\
In the present work, for the first time, a systematic design approach for backbone tracking using a phase-locked loop is proposed.
The goal is to develop an approach that is widely applicable, easy to implement, and requires minimal prior knowledge of the system.
%An important requirement is that the approach is sufficiently general that it is useful for a wide range of systems, and th
%All parameters, including the gains of the proposed proportional-integral controller, and the parameters of the adaptive filter, are designed, relying on data acquired via conventional shaker-based linear modal testing, and an open-loop sine test at a single frequency and level.
In \sref{problemSetting}, the problem setting is formulated.
The theory behind the proposed solution approach is described in \sref{proposition}.
An algorithmic summary of the proposed approach and practical recommendations are given in \sref{practical}.
Numerical validation and an experimental assessment are presented in \sref{numerical} and \sref{experimental}, respectively.
Concluding remarks are made in \sref{conclusions}.

%%%%%%%%%%%%%%%%%%%%%%%%%%%%%%%%%%%%%%%%%%%%%%%%%%%%%
\section{Problem setting\label{s:problemSetting}}
% PURPOSE OF SECTION
The problem setting is schematically illustrated in \fref{PLL}.
The closed loop consists mainly of a plant, a phase detector, and a controller.
The plant is the union of the structure under test and the vibration exciter.
An excitation via a single electro-dynamic shaker with a stinger is considered (\fref{plant}), which is the by far most popular setup for nonlinear vibration testing.
The transfer of the proposed approach to base excitation seems feasible \cite{Muller.2022}, but is viewed as outside the scope of the present work.
The applied force $f$ and the response $\qex$ are measured.
As stated above, an adaptive filter is used for phase detection and, more generally, for Fourier decomposition.
The difference of the estimated phase lag to the set value (reference) is fed to the controller, which outputs a frequency.
%At this point, the \emph{phase} is said to have \emph{locked}, the duration until this point is called \emph{(phase) settling time}.
Integration of the frequency yields the phase, which is used as argument of a harmonic function that modulates the voltage input to the exciter.
In the following, the mathematical model of the closed loop is formulated, fundamental assumptions are specified, and the design parameters to be set by the proposed approach are identified.
\begin{figure}[t]
    \centering
    \includegraphics[width=\textwidth]{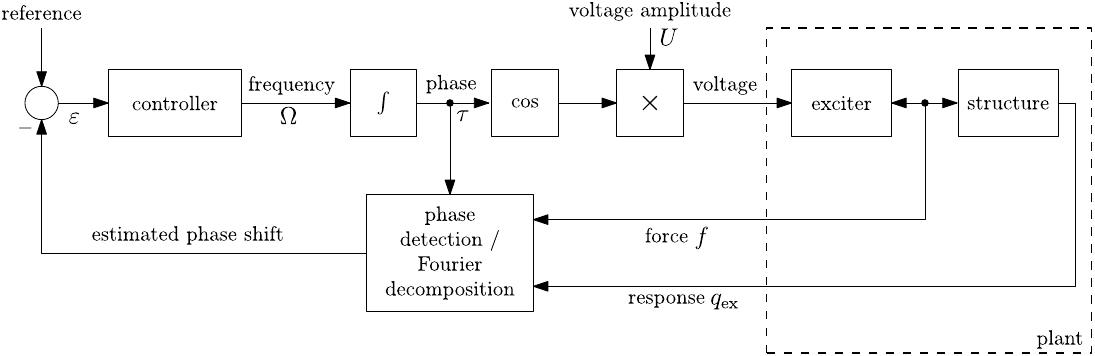}
    \caption{Schematic of considered problem setting.}
    \label{f:PLL}
\end{figure}
\begin{figure}[t]
    \centering
    \includegraphics{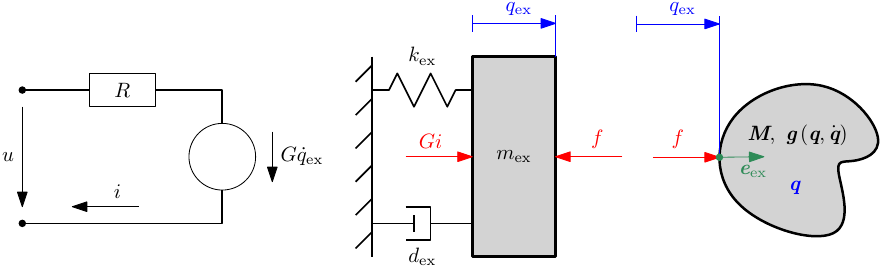}
    \caption{Model of the plant: (left) electrical and (middle) mechanical part of the exciter; (right) structure under test.}
    \label{f:plant}
\end{figure}
\\
% STRUCTURE UNDER TEST
In most previous works, a Duffing oscillator was considered as plant model \cite{Fan.2007,Denis.2018}.
For wider applicability, a more generic model of the structure under test is considered,
\ea{
\mm M \ddot{\mm q} + \mm g\left(\mm q,\dot{\mm q}\right) &=& \eex f\fp \label{e:SUT}
}
Herein, $\mm q\in\mathbb R^{n\times 1}$ is the vector of generalized coordinates, where $n$ is the number of degrees of freedom, overdot denotes derivative with respect to time $t$, $\mm M=\mm M^{\mrm T}>\mm 0$, $\mm M\in\mathbb R^{n\times n}$ is the mass matrix, $\mm g$ are generalized forces, and $\eex$ describes point and direction of the applied force $f$.
%(linear and nonlinear, damping and restoring forces)
The limiting assumptions on the behavior of the structure are adopted from those underlying the Extended Periodic Motion Concept \cite{Krack.2015a}:
$\mm q=\mm 0$ is an asymptotically stable equilibrium position, and the frequency of the target mode must be away from any internal resonance condition.
%The latter means that the corresponding modal frequency must be well-separated, and the structure must be sufficiently far away from internal resonance.
This non-resonance condition must hold not only in the linear case, but up to the vibration level of interest.
Under those conditions, the structure behaves like a single nonlinear modal oscillator, so that one can achieve phase resonance with a single exciter, and restricting the control to the fundamental harmonic \cite{Scheel.2018,Muller.2022}
\\
% EXCITER
The goal is commonly to identify the dynamic behavior of the structure under test, not of the plant (which also includes the exciter).
To make this distinction, the behavior of the exciter has to be considered.
%In the proposed approach, in particular, the damping provided by the exciter is taken into account.
In the present work, the conventional model of an electro-dynamic exciter is used \cite{Tomlinson.1979,McConnell.2008},
\ea{
\mex\ddqex + \dex\dqex + \kex\qex &=& G i - f\fk \label{e:exciterEoM}\\
Ri + G\dqex &=&  u\fk \label{e:exciterEL}\\
\qex &=& \eex^{\mrm T}\mm q\fp \label{e:compatibility} % equilibrium (actio=reactio) implicitly considered
}
It is assumed that the stinger is rigid, so that the shaker armature is regarded as directly attached to the structure at the drive point (\eref{compatibility}).
In \eref{exciterEoM}, $\mex$ is the moving (or dynamic) mass of the exciter.
This includes the armature, the coil, possible sensors and the stinger.
$\kex$, $\dex$ denote the mechanical stiffness and damping of the exciter, $G>0$ is the electromotive force constant, % = inductance/back-EMF constant? --> check previous papers
$R>0$ is the electrical resistance,
$i$ is the current,
$u$ is the voltage.
The self-induction within the electrical part of the exciter is neglected, which is a very common assumption at sufficiently low frequencies.
Physically, this means that there is no phase lag between the voltage and the voltage-imposed part of the current (excluding the term $G\dqex$ in \eref{exciterEL}).
The main intentions behind this assumption were to avoid more lengthy expressions in the theoretical derivation, and to avoid having to identify this parameter.
In fact, the proposed control design will turn out to require only very few system parameters as input, which are to be obtained based on a conventional shaker-based linear modal analysis, as explained in \aref{exciterParameters}.
%The model of the structure under test is formally introduced here for the mathematical derivation of proposed control design scheme.
%Of course, such a model is not always available in practice; in fact, the purpose of the phase resonance test is to identify the dynamic behavior of the structure.
%linear frequency response test around target mode %; such a test is commonly done and highly recommended in any case before a nonlinear test % OR after $\omex$ and $\Dex$ are introduced in the slow dynamics section?
\\
% phase detector / FOURIER DECOMPOSITION
The proposed adaptive filter is governed by the set of ordinary differential equations\footnote{It should be remarked that adaptive filters are more commonly stated in time discrete form; the time-continuous form in \erefs{AFF}-\erefo{AFQ} is derived in \aref{adaptiveFilter}.},
\ea{
\dot{\hat{F}}^{(h)} &=& 2\omLP \ee^{-\ii h\tau}\left(f-\real{\sum\limits_{h=0}^{H} \ee^{\ii h\tau}\hat{F}^{(h)}}\right) \quad h=0,\ldots,H\fk\label{e:AFF}\\ % ASK PATRICK TO CHECK CORRECNTESS OF SIGNS OF COMPLEX EXPONENTIALS THROUGHOUT PAPER
\dot{\hat{Q}}^{(h)} &=& 2\omLP \ee^{-\ii h\tau}\left(\qex-\real{\sum\limits_{h=0}^{H} \ee^{\ii h\tau}\hat{Q}^{(h)}}\right) \quad h=0,\ldots,H\fp \label{e:AFQ}
}
Herein, ${\hat{F}}^{(h)}$ is the estimate of $h$-th complex Fourier coefficient of $f$, and ${\hat{Q}}^{(h)}$ is the estimate of the $h$-th complex Fourier coefficient of $\qex$.
At steady state, the truncated Fourier series in the parenthesis of \eref{AFF} should be a very good approximation of $f$, implying $\dot{\hat{F}}^{(h)}\approx\mm0$.
The filter parameters are the order $H$, and the coefficient $\omLP>0$.
The interpretation of $\omLP$ as cutoff frequency of a low pass filter is established in \ssref{slow}.
$\tau$ is the integral of the instantaneous frequency $\Omega$ (\eref{tau}), which, in turn, is the output of the controller (\eref{PI}).
Thus, $\Omega$ is generally time-variable but known, which is in contrast to most applications of adaptive filters, where constant-frequency terms are considered.
\\
% VCO, controller, amplifiers, sensors
The remaining equations of the closed loop system are:
% (goverened by the whole of \erefs{SUT}-\erefo{thest})
%
\ea{
u &=& \Uc\cos\tau\fk \label{e:udef}\\
\tau &=& \int_0^t\Omega\dd \check t\fk \label{e:tau}\\
\Omega &=& \Omega_{\mrm{ini}} + \kp\err+\ki I_{\err}\fk \label{e:PI}\\
\err &=& \thref - \left(\thfest-\thest\right)\fk \label{e:err}\\
\dot I_{\err} &=& \err\fk \label{e:Ierr}\\
\thfest &=& \Arg{\hat F^{(1)}}\fk \label{e:thfest}\\
\thest &=& \Arg{\hat Q^{(1)}}\fp \label{e:thest}
}
%
% Amplifier/Gain
Herein, $\Uc>0$ is the voltage amplitude.
$\Uc$ is prescribed in the present work; the exciter is operated in voltage mode.
The addition of an amplitude controller is viewed as a natural extension for future work.
% PI controller
A simple proportional-integral phase controller is used in the present work.
Compared to a pure integral controller (gain $\ki$), introducing a proportional term (gain $\kp$) is known to enable quicker settling.
An additional differential term could be used, but it would increase the sensitivity to noise, and it will be shown that noise is the dominant impediment when targeting speed and robustness.
$I_{\err}$ is an auxiliary state variable.
$\err$ is the control error.
It is defined as the deviation from local phase resonance between fundamental harmonics of forcing and response displacement.
If the velocity or the acceleration is measured instead of the displacement, the reference phase $\thref$ has to be adjusted accordingly.
The control error is obtained from the current estimates, $\thfest$ and $\thest$, of the phase of the force and the response displacement, which are determined from the output of the adaptive filter (\erefs{thfest}-\erefo{thest}).
\\
% DESIGN VARIABLES|PARAMETERS
In summary, for a given plant configuration (structure under test with attached exciter), the parameters of the closed loop are:
\begin{itemize}
    \item initial frequency: $\Omega_{\mrm{ini}}$\fk
    \item voltage amplitude: $\Uc$\fk
    \item cutoff frequency and order of adaptive filter: $\omLP$, $H$\fk
    \item control gains: $\kp$, $\ki$\fp
\end{itemize}
The purpose of the present work is to design these parameters to robustly and quickly track the phase resonant backbone curve.
Some of the parameters are easier to select than others:
The initial frequency $\Omega_{\mrm{ini}}$ should simply be near the (linear) modal frequency of the target mode.
The voltage amplitude $\Uc$ is stepped to reach different vibration levels.
The order of the filter $H$ has to be as high as necessary to capture all relevant harmonics of force and response.
Most of the subsequently established theory addresses the selection of the control gains $\kp$, $\ki$ and the cutoff frequency $\omLP$, with the goal to robustly and quickly reach a phase-locked state.
The theory underlying the proposed approach is described in \sref{proposition}.
Practical recommendations along with a summary are given in \sref{practical}.

\section{Theory behind proposed approach\label{s:proposition}}
In this section, the theory behind the proposed approach for robust and quick backbone tracking is established, for the generic plant model (\erefs{SUT}-\erefo{compatibility}) with the controller defined by \erefs{AFF}-\erefo{thest}.
A fundamental assumption is that the amplitude and the phase lag of the plant \emph{evolve slowly} as compared to the \emph{fast oscillation} with frequency $\Omega$, under the action of the phase controller.
This implies that $\Omega$ changes slowly.
The slow-fast decomposition permits to extract the dynamics on the slow time scale using an averaging formalism (\ssref{slow}), which leads to a set of autonomous first-order ordinary differential equations.
The fixed point of those equations corresponds to the phase-locked state (\ssref{locked}).
The asymptotic behavior around the locked state is then analyzed using linear theory (\ssref{asympt}).
By making a few simplifying assumptions (phase-neutral exciter, structure operated in linear regime), the problem becomes amenable to analytical solution.
It is shown that the phase transient evolves on the time scale induced by the cutoff frequency of the adaptive filter; \ie, the higher $\omLP$, the faster the phase locks.
The control gains $\kp$, $\ki$ are selected to optimize the settling of the phase transient.

\subsection{Dynamics on slow time scale\label{s:slow}}
The dynamics on the fast time scale, \ie, the oscillation with frequency $\Omega$, is induced by \erefs{udef}-\erefo{tau}.
It will be assumed that $\Omega>0$.
In principle, $\Omega\leq0$ is possible in the case of very poor initialization, \eg, $\Omega_{\mrm{ini}}<0$, or in the case of a divergent controller (cf. \eref{PI}).
Hence, assuming $\Omega>0$ is not an important restriction in practice.
To extract the dynamics on the slow time scale, an averaging formalism is applied.
This consists in taking the integral over one period (related to $\Omega$), treating the slowly varying quantities as time-constant.
This eliminates the fast time scale $\tau$.
%\\
Averaging is first applied to the plant in \sssref{slowPlant}, and then to the adaptive filter in \sssref{slowFilter}.
The remaining equations of the closed loop system are either algebraic or already in first-order form, so that averaging yields trivial results\footnote{The structure of the equations does not change. Formally, the quantities have to be replaced by their mean values.}.

\subsubsection{Plant dynamics on slow time scale\label{s:slowPlant}}
% IDEA
In this section, the model of the structure (\eref{SUT}) is reduced to a \emph{single nonlinear modal oscillator}.
This greatly simplifies the subsequent derivations.
In contrast to previous works that analyze the stability of phase-locked loops for vibration testing \cite{Sun.2002,Fan.2007,Denis.2018}, no restriction to a certain type of nonlinearity is necessary, since the modal properties offer a non-parametric description of rather generic nonlinear terms of arbitrary physical origin.
However, the derivation relies on the assumption that such a single-mode reduction is possible.
%[an important|the] purpose of the backbone tracking is to identify the amplitude-dependent modal frequency and damping ratio, this reduction seems natural / does not introduce further restrictions%- as will be shown, this allows us to justify local, fundamental phase resonant forcing; and
%- this permits to express modal properties (both frequency and damping) as function of directly measured quantities, in closed form
%\\
% EXTENDED PERIODIC MOTION CONCEPT
In particular, as stated in \sref{problemSetting}, we assume that the equilibrium position $\mm q=\mm 0$ is asymptotically stable, and that the target modal frequency is well-separated and not in internal resonance.
Under those conditions, a single nonlinear mode dominates the response around the phase-resonant backbone curve.
Here, a nonlinear mode is defined according to the Extended Periodic Motion Concept \cite{Krack.2015a}, as a family of periodic solutions of $\mm M\ddot{\mm q} + \mm g\left(\mm q,\dot{\mm q}\right) - 2D\omega\mm M\dot{\mm q} = \mm 0$, which continues the target linear mode of the linearized system from $\mm q=\mm 0$ to finite vibration levels.
The artificial negative damping term $2D\omega\mm M\dot{\mm q}$ cancels the natural dissipation in period-average.
$\omega>0$ is the fundamental (angular) modal frequency of the periodic oscillation, $D$ is modal damping ratio.
%Both depend on vibration level.
The periodic modal oscillation is decomposed into a Fourier series, where the complex Fourier coefficients are expressed as $a \mm\phi^{(h)}$, and $a>0$ denotes the modal amplitude.
Mass-normalization is used so that $\left(\mm\phi^{(1)}\right)^{\mrm H}\mm M\mm\phi^{(1)} = 1$, where $\square^{\mrm H}$ denotes the complex conjugate transpose (Hermitian).
It should be emphasized that the Fourier coefficients of the modal oscillation are generally complex, allowing for non-trivial phase lags among the material points of the structure under test. % (typical for moderate local damping).
The modal properties $\omega$, $D$, $\mm\phi^{(h)}$ depend on $a$, which is not explicitly denoted for brevity.
%- where needed, the corresponding asymptotic values (linear regime) are explicitly indicated as $\omega_{\mrm{lin}}$, $D_{\mrm{lin}}$ % $=\omega(a\to0)$, $=D(a\to0)$
% - artificial negative damping term $2D\omega\mm M\dot{\mm q}$: consistent with conservative case ($D=0$), and with linear case of modal damping (thanks to orthogonality of modes w.r.t. this term); otherwise, \ie, for high damping and strong contribution of multiple modes, this term may distort the natural dynamics in an intrusive way that the modal properties become meaningless / in particular, non-representative of the near-resonant dynamics; this limitation is not important in the present work, as we have to assume that strong modal interactions are absent to apply SNMT
\\
% SINGLE NONLINEAR MODE THEORY
To derive the governing equation of the single nonlinear modal oscillator, the approximation $\mm q\approx \mathcal U(a,\th)$ is made with
\ea{
\mathcal U(a,\th) &=& \real{\sum\limits_{h=0}^{\infty} a\mm\phi^{(h)}\ee^{\ii h\left(\tau+\th\right)}}\fp \label{e:qansatz} % \\
%\dot{\mm q} \approx \mathcal V(a,\th) &=&  \real{\sum\limits_{h=0}^{H} (\ii n\Omega)a\mm\phi^{(h)}\ee^{\ii n\left(\tau+\th\right)}}\fp \label{eq:e:vansatz}
}
\eref{qansatz} is substituted into \eref{SUT}, and it is required that the residual is orthogonal with respect to the fundamental harmonic of the mode.
Here, amplitude $a$ and phase lag $\th$ are allowed to vary slowly with time, so that averaging can be applied.
This way, one obtains \cite{krac2014a}
\ea{
2\ii\Omega\left(\dot a + \ii a \dth\right) + \left( -\Omega^2 + 2D\omega\ii\Omega + \omega^2 \right)a &=& \mm\phi^{\mrm H}\eex F\ee^{-\ii\th}\fp \label{e:SNMT}
}
Here and in the following, the abbreviation $\mm\phi=\mm\phi^{(1)}$ is used, and similarly $F=F^{(1)}$.
% higher harmonics will turn out not to be relevant on the slow time scale
Requiring orthogonality only with respect to the fundamental harmonic permits to express the nonlinear term in the projected equation \erefo{SNMT} in closed form, using the amplitude-dependent modal properties.
% slight inaccuracy when higher harmonics become prominent
% not fully consistent in the Galerkin sense
\\
The right-hand side of \eref{SNMT} is obtained by considering \erefs{exciterEoM}-\erefo{udef}:
\ea{
\mm\phi^{\mrm H}\eex F &=& \frac{\phex G \Uc}{R} - \left( -\Omega^2 + 2 \ii \Omega \Dex\omex + \omex^2 \right)  \muex a \ee^{\ii\th}\fp \label{e:F}
}
Herein, $\omex$ and $\Dex$ are the natural frequency and the damping ratio of the mechanical part of the exciter, where $\omex = \sqrt{\kex/\mex}$, and $2\Dex\omex=(\dex+G^2/R)/\mex$.
Note that the back-electromotive force leads to an apparent viscous damping; \ie, $G^2/R$ is added to the mechanical damping $\dex$.
Further, $\phex = \mm\phi^{\mrm H}\eex$, and $\muex = \phex^2\mex$ denotes the modal mass ratio.
For convenience, the phase is normalized in such a way that $\phex\in\mathbb R$ and $\phex>0$.
This can be done without loss of generality, provided that the shaker is not attached at a vibration node.
%
% PROBABLY UNNECESSARY DETAIL:
%- note that thanks to assumed rigid attachment of exciter mass to structure, $\qex$ is not a separate degree of freedom, so that we do not have to apply averaging separately to \eref{exciterEoM}
%- also note that electrical part of model is algebraic (no derivative of electrical current $i$ appears in the governing equations), so that no averaging here either

\subsubsection{Adaptive filter dynamics on slow time scale\label{s:slowFilter}}
By applying averaging to both sides of \eref{AFF} and \eref{AFQ}, \ie taking the integral over one period, one obtains for $h=1$:
\ea{
\dot{\hat{F}} &=& \omLP \left(F-\hat F\right)\fk \label{e:AFFavg}\\
\dot{\hat{Q}} &=& \omLP \left(\phex a\ee^{\ii\th}-\hat Q\right)\fp \label{e:AFQavg}
}
%
%Herein, $\dot{\hat{F}}$ is period-average of $\dot{\hat{F}}^{(1)}$, and analogous for $\hat Q$
It is important to note that \erefs{AFFavg}-\erefo{AFQavg} are independent of the higher harmonics, thanks to the pairwise orthogonality of the harmonic functions $\ee^{\ii h\tau}$ over a period.
This also applies to the averaged filter equations corresponding to higher harmonics ($\hat F^{(h)}$ with $h\neq 1$ and analogous for $\hat Q^{(h)}$); \ie, the averaged filter equations are harmonically decoupled.
Further, the interpretation of $\omLP$ is now evident from \erefs{AFFavg}-\erefo{AFQavg}:
In period-average, the adaptive filter acts as first-order low-pass filter with the cutoff frequency $\omLP$.
\\
% EFFECT OF \omLP
The higher $\omLP$, the faster will the filter reduce the deviation between the Fourier coefficient of the input, $F$, and the estimate $\hat F$.
From the perspective of the slow time scale, thus, $\omLP\to\infty$ seems best.
However, higher $\omLP$ lead to more pronounced fluctuation of $\hat F^{(1)}$ on the fast time scale, as explained later. % (\ssref{omLP}).
Indeed, the dominant constraint for maximizing $\omLP$ is noise.
Real noise can have various sources and does not have to be a stationary Gaussian process.
In lack of a universally valid noise model, a practical approach for the selection of a suitable $\omLP$ is proposed in \ssref{omLP}.

\subsection{Fixed point on slow time scale: the locked state\label{s:locked}}
% ODE IN WORDS
The dynamics of the closed loop on the slow time scale can be expressed as a system of explicit first-order ordinary differential equations. %; for constant parameters ($\omLP$, $\kp$, $\ki$, $\Uc$) % $\Omega_{\mrm{ini}}$, $\thref$
The state variables are (the period-average of) $a$, $\th$, $\hat F$, $\hat Q$, and $I_{\err}$, which are governed by the ordinary differential equations \erefo{SNMT}, \erefo{AFFavg}-\erefo{AFQavg}, and \erefo{Ierr}.
The equation system is closed by the algebraic equations \erefo{PI}-\erefo{err}, \erefo{thfest}-\erefo{thest}, and \erefo{F}.
\\
% FIXED POINT
At the \emph{fixed point}, the state variables are time-constant, which yields the algebraic relations:
\ea{
0 &=& \err\fk \\
I_{\err} &=& \frac{\Omega_{\mrm{ini}}-\Omega}{\ki} \fk \\
0 &=& \frac{\pi}2 - \left(\thfest - \hat{\th}\right)\fk \\
\hat F &=& F = |F|\ee^{\ii\thf}\fk\\
\hat Q &=& \phex a\ee^{\ii\th}\fk\\
\thest &=& \th\fk\label{e:thestth}\\
\thfest &=& \thf\fk\label{e:thfestthf}\\
\Omega &=& \omega\fk \label{e:Omom}\\
2D\omega^2a &=& \phex |F|\fp
}
%
%In practice, one can recognize that the fixed point is reached, by observing that the period average of $\Omega$, $\err$, $I_{\err}$ and the estimated Fourier coefficients has settled.
When the control error $\err$ vanishes in period average, the period average of $I_{\err}$ is constant. % ($I_{\err}\neq0$ in general)
It follows that the period-average of $\Omega$ is also constant, and equals to the modal frequency, $\Omega=\omega$.
This implies that the closed loop has locked onto the phase resonant backbone of the structure under test.

\subsection{Design of asymptotic behavior around the locked state\label{s:asympt}}
% IDEA/LIMITATIONS
Ideally, the locked state is reached rapidly from arbitrary initial conditions.
Global stability analysis would require complete knowledge of the nonlinear system, which is generally not available.
%(which we do not have at the design stage of the controller)
Instead, the asymptotic behavior around the locked state will be designed.
%- local knowledge becomes available during testing (in terms of $D$, $\omega$, $\phex$ as function of amplitude)
%- amplitude-adaptive control design is considered for future work
%- present work: design controller for an amplitude in (almost) linear regime, and see if constant controller parameters lead to well-performing behavior also at higher amplitudes; important benefit: simplicity (fixed parameters)
%- in fact, with a further simplifying assumption (phase-neutral exciter), an [analytical|closed-form] design is possible; clear understanding of parameter dependencies
\\
% LINEAR ODE
As shown in \aref{derivation}, one can reduce the state-space dimension in the linear case.
More specifically, one can replace the complex state variables $\hat F$, $\hat Q$ by the control error $\err$.
Further, the normalized time $\bar{t}=\omLP t$ is introduced, and the state $\bar I_{\err} = \omLP I_{\err}$ is used instead of $I_{\err}$.
Thus, the new state vector is $\mm z = [a;\th;\err;\bar{I}_{\err}]$ (semicolon denotes vertical concatenation).
First-order Taylor series expansion around the fixed point yields a linear autonomous ordinary differential equation system.
The corresponding coefficient matrix contains the modal properties and their derivatives with respect to $a$, evaluated at the fixed point.
%This could be the point of departure for a control design with amplitude-adaptive parameters in the future.
In the present work, it is proposed to design the control parameters for the low-level regime of the structure under test, and to use these constant parameters throughout the backbone.
%for the complete backbone curve.
%More specifically, the control parameters are designed for the low-level regime of the structure under test, where the nonlinear modal properties can be replaced by their linear counterparts.
In addition, a phase-neutral exciter is assumed, which further simplifies the problem, so that an analytical solution can be obtained, enabling a clear understanding of the essential parameter dependencies.
\\
% FURTHER SIMPLIFIED ODE
With the above stated simplifications, the system of linear autonomous ordinary differential equations is (\aref{derivation}):
\ea{
\Delta \mm z^\prime &=& \mm A_0 \Delta \mm z\fk \label{e:dz}\\
\mm A_0 &=& \matrix{cccc}{
-\delplnd & 0 & -\kpnd\Upsilon & -\kind\Upsilon\\
0 & -\delplnd & -\kpnd & -\kind\\
0 & \frac{\delpl}{\dellin} & -1 & 0\\
0 & 0 & 1 & 0
}\fk \label{e:A0}
}
where $\Delta\mm z$ is the deviation from the considered fixed point, $\square^\prime=\dd\square/\dd\bar t$, and
\ea{
\dellin &=& \left(D\omega\right)_{\mrm{lin}}\fk\\
\delpl &=& \dellin + \left(\muex\Dex\omex\right)_{\mrm{lin}}\fk\label{e:delpl}\\
\delplnd &=& \frac{\delpl}{\omLP}\fk\\
\kpnd &=& \kp\frac{\left(1+\muex\right)_{\mrm{lin}}}{\omLP}\fk\label{e:kpnd}\\
\kind &=& \ki\frac{\left(1+\muex\right)_{\mrm{lin}}}{\omLP^2}\fk\label{e:kind}\\
\Upsilon &=& \frac{GU}{2R}\left(\frac{\phex}{(1+\muex)\omega^2}\right)_{\mrm{lin}}\fp
}
In the following, positive damping is assumed, of both the structure ($\dellin>0$) and the plant ($\delpl>0$).
\\
% AMPLITUDE TRANSIENT
In the first column of $\mm A_0$ in \eref{A0}, only the first element is nonzero.
This means that the evolution of the amplitude does not affect the phase lag, nor the phase control error (or its integral in time).
This is an important simplification. %, and this was obtained for the assumption that the amplitude-dependence of the modal properties is small, and the exciter is in good approximation phase-neutral
One can see that $\mm A_0(1,1)=-\delplnd$ is an eigenvalue of $\mm A_0$, associated with the \emph{amplitude transient}.
%eigenvector $[1;0;0;0]$ (because the lower $3\times 3$ sub-matrix of $\mm A_0$ has full rank (presuming $\delpl,\ki\neq0$))
%- this represents the \emph{amplitude transient}, which
Thus, the amplitude transient decays exponentially with $\ee^{-\delplnd \bar t}$ in the neighborhood of the locked state.
From \eref{delpl}, it can be found that the damping provided by the exciter actually has a positive effect on the decay of the amplitude transient ($\delpl>\dellin$).
As discussed below \eref{F}, the exciter damping is composed of a mechanical damping and a back-electromotive force.
In the current mode of operation, the exciter amplifier tries to mitigate the back-electromotive force.
Thus, for very lightly damped structures, it can be beneficial to operate the exciter amplifier in the \emph{voltage mode} (as presumed throughout this work).
% AMPLITUDE CONTROL
%If the amplitude transient takes too long, one should apply response amplitude control, as discussed later in \ssref{summary}
It is useful to note that the amplitude transient is governed by the plant, and thus the phase controller has no influence on it.
For very lightly-damped structures, it is likely that the phase settles before the amplitude, especially with the optimized phase transient proposed in this section.
Thus, one effectively sweeps along the backbone.
%Given that the adaptive filter output is trustworthy, one thus obtains continuous backbone data.
If the amplitude transient is deemed too slow, one should extend the feedback loop by a response amplitude controller.
%It is common to combine phase with amplitude control. %(forcing \cite{Peter.2017,Schwarz.2020}; response \cite{Woiwode.2024}).
\\
% PHASE TRANSIENT
The \emph{phase transient} is determined by the eigenvalues of the lower $3\times 3$ sub-matrix of $\mm A_0$.
To have asymptotic stability, all eigenvalues must have negative real part.
The quickest decay of the phase transient is expected when the maximum real part among the eigenvalues is minimized.
To achieve this, it is useful to note that the trace of this sub-matrix equals the sum of the eigenvalues,
\ea{
\lambda_1 + \lambda_2 + \lambda_3 = -\left(\delplnd+1\right)\fp
}
Since $\Delta\mm z$ and $\mm A_0$ are real, we generally have either three real, or one real and a pair of complex-conjugate eigenvalues.
In either case, the sum is real, and $-\left(\delplnd+1\right)/3$ is the mean value of the real parts of the eigenvalues.
Hence, to minimize the maximum real part among the eigenvalues, $\lambda_1$, $\lambda_2$ and $\lambda_3$ should all have the same real part
\ea{
\lamRnd = -\frac{\delplnd+1}3\fp \label{e:lamRopt}
}
%if any real part is shifted to the left of this in the complex plane, there will be at least one eigenvalue with real part larger than this;
With this, the desired eigenvalues can be expressed as
\ea{
\lambda_1 &=& \lamRnd\fk\label{e:lam1}\\
\lambda_2 &=& \lamRnd + \ii \lamInd\fk\label{e:lam2}\\
\lambda_3 &=& \lamRnd - \ii \lamInd\fk\label{e:lam3}
}
where $\lamInd\geq0$ without loss of generality.
\\
As shown in \aref{derivation}, the above eigenvalue setting can be achieved by selecting the control gains as follows:
\ea{
\kpnd &=& \frac{\dellin}{\delpl} \left( 3\lamRnd^2 + \lamInd^2 - \delplnd \right)\fk \label{e:kp_gen}\\
\kind &=& -\frac{\dellin}{\delpl} \lamR \left( \lamRnd^2 + \lamInd^2 \right)\fp \label{e:ki_gen}
}
It is easy to see that $\kind>0$ since $\lamR<0$.
By substituting $\lamRnd$ from \eref{lamRopt} into \eref{kp_gen}, one can also establish that $\kpnd>0$ (for the assumed positive damping). %(for $\delpl>0$)
\\
% IMAGINARY PART ZERO IS NOT A GOOD IDEA
An appropriate value for the imaginary part $\lamInd$ still needs to be chosen.
It is tempting to simply choose $\lamInd=0$, as this should lead to the least-oscillatory phase transient.
However, this leads to sub-optimal performance:
Indeed, from \erefs{kp_gen}-\erefo{ki_gen}, one can follow that $\lamInd=0$ leads to the smallest control gains, which has the tendency to cause large phase errors during the transient.
%Also, any imperfection (\eg nonlinearity, time-variability, exciter not phase neutral), is likely to lead to a situation where one of the eigenvalues has a different real part than the others.
%To find a suitable value for $\lamInd$, it is proposed to design the phase transient representative of a step along the backbone, as explained in the following.
In the following, a more appropriate choice of $\lamInd$ (with $\lamInd>0$) is proposed, for the given task of backbone tracking.
\begin{figure}[htb]
    \centering
    \begin{subfigure}{0.49\textwidth}
    \centering
        \includegraphics[]{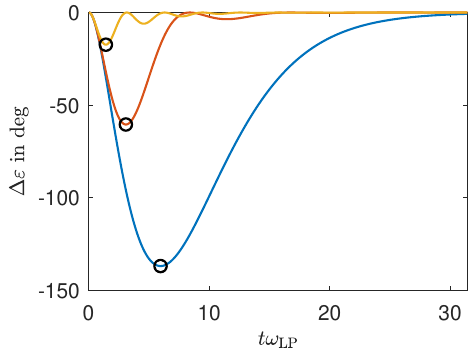}
        \caption{}
        \label{f:opti_eps}
    \end{subfigure}
    \hfill
    \begin{subfigure}{0.49\textwidth}
    \centering
        \includegraphics[]{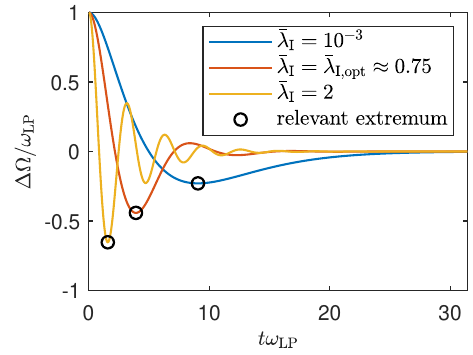}
        \caption{}
        \label{f:opti_Om}
    \end{subfigure}
    \caption{Phase error (a) and frequency shift (b) over time for a unit initial frequency offset. The point labeled \textit{relevant extremum} is used to define the optimal $\lamInd$. $\delplnd = 10^{-2}$, $\muex=0$ (implying $\delpl/\dellin=1$).
    }
    \label{f:optiExample}
\end{figure}
\begin{figure}[!ht]
    \centering
    \includegraphics[]{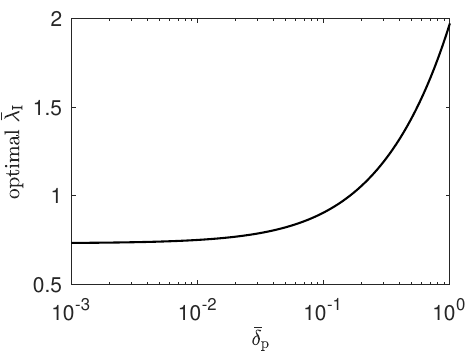}
    \caption{Optimum $\lamInd$ leading to the proposed trade-off between control error and frequency overshoot during a step along the backbone. $\delplnd$ is the normalized decay rate.}
    \label{f:solOpti}
\end{figure}
\\
Recall that the goal is to track the backbone curve, which is implemented by successive stepping of the voltage amplitude.
Suppose that we have reached a steady state, and then increase the excitation level.
We thus start from phase resonance ($\err=0=\Delta\err$, $\Delta\th = 0$), and asymptotically approach this condition afterwards.
In the general case, however, the modal frequency will be different, which corresponds to a nonzero initial value for $\Delta\bar{I}_{\err}$.
In \fref{optiExample}, the solution of the corresponding initial value problem (phase transient) is illustrated for different settings of $\lamInd$.
More specifically, the evolution of the (phase) control error, $\Delta\err$, and the frequency, $\Delta\Omega$, is determined by solving \eref{dz} for the described initial values. %$\Delta\mm z = [0;0;0;1]$
Clearly, the higher $\lamInd$, the smaller the maximum control errors, but the larger the frequency overshoots initially.
Interestingly, control error and frequency overshoot reach finite limit values for $\lamInd\to0$ and $\lamInd\to\infty$, respectively.
As a trade-off, we propose to determine $\lamInd$ so that maximum control error and maximum frequency overshoot, normalized by their respective limit values, are minimized.
While the initial value problem can be solved analytically, this condition leads to a transcendental equation, as shown in \aref{derivation}, which has to be solved numerically.
%The solution depends only on the parameter $\delplnd$.
The solution of that equation gives a unique trade-off $\lamInd$ for each $\delplnd$, as illustrated in \fref{solOpti}.
From this figure, one can infer that the higher the plant damping, the higher the proposed $\lamInd$, and, in turn, the higher the control gains.
It will later be seen, however, that $\delplnd<10^{-1}$ for reasonably fast adaptive filters and light damping, so that the optimal $\lamInd$ does not deviate much from its asymptotic value reached for $\delplnd\to 0$.
% NOTE: phase error in degrees is huge, but this does not matter, of course, since the problem is linear
% For the interpretation of this figure, it is useful to recall $\Delta\Omega/\omega = \Delta\Omega/\omega*\omega/\omLP$ where $\omega/\omLP$ is typically in $[0.01,0.1]$.
%Also, it is interesting to note that $\Delta\Omega/\omLP = \bar{\kp}\Delta\err + \bar{\ki}\Delta\bar{I}_{\err}$

\section{Practical recommendations and summary of proposed approach\label{s:practical}}
Recall that for a given plant configuration (structure under test with attached exciter), the parameters of the phase resonance test are: initial frequency $\Omega_{\mrm{ini}}$, voltage amplitude $\Uc$, cutoff frequency $\omLP$ and order $H$ of the adaptive filter, as well as the control gains $\kp$, $\ki$.
It is proposed to set $\Omega_{\mrm{ini}}=\omlin$, where $\omlin$ is identified using a conventional linear modal test.
The setting of the limits and increments of the voltage amplitude $\Uc$ is discussed in \ssref{voltageRamps}, and it is proposed to increase the voltage along ramps during the backbone tracking.
Theoretical considerations on $\omLP$ and $H$ were discussed in \sssref{slowFilter}; a practical, yet systematic tuning procedure is proposed in \ssref{omLP}.
Optimal values for the control gains were derived in \ssref{asympt}.
Recommendations on the automatic settling detection and the elimination of hold times are made in \ssref{lockInDetect} and \sref{removeHoldTimes}, respectively.
Finally, the overall approach is summarized in an algorithmic form in \ssref{summary}.
As the purpose of backbone tracking is often the identification of the amplitude-dependent modal properties, these are expressed as function of directly acquired quantities in \ssref{mprop} for completeness.

\subsection{Voltage ramps\label{s:voltageRamps}}
To track the backbone curve, the vibration level needs to be adjusted.
Usually one would like to acquire data also in the almost linear regime, in order to check consistency with results obtained from conventional linear modal testing.
A reasonable signal-to-noise ratio is required to obtain a sufficiently robust and fast adaptive filter.
This can lead to a lower bound of the vibration level in practice, as discussed below (\ssref{omLP}).
The upper bound of the vibration level will be application specific.
Examples are the exciter limitations, the intent to avoid structural damage, or the expected vibration level under operating conditions.
A sequence of voltage amplitudes $\lbrace{U_1,\ldots,U_{N_u}}\rbrace$ should be specified by the user.
The step size should be as small as necessary for the linear theory underlying the control design to hold (\ie, the system remains close to phase resonance / the phase control error remains small).
Also, it should be as small as necessary to capture the amplitude-dependence of the quantities of interest with sufficient resolution.
On the other hand, the step size should be as large as possible to avoid spurious effort.
\\
The common step-wise (discontinuous) change of $\Uc$ from one specified value to the next could introduce sudden changes in the response.
In particular, it could cause higher harmonic and modal distortions.
%es, which could distort the dynamics of interest. %(with unpredictable outcome) % relatively long transients; violate the slow-fast decomposition underlying the proposed control design approach
For a smoother transition between voltage levels, half-cosine ramps are proposed,
\ea{
U(t) = U_{i-1} + \frac{1}{2} \left[ 1 - \cos \left(\frac{2\pi}{T_\mathrm{ramp}}\left(t-t_{i}\right)\right)\right] \left(U_{{i}} - U_{i-1}\right), \qquad t_{i} \leq t \leq t_{i} + T_\mathrm{ramp}\fk\label{e:Uramp}
}
Where for $i=1$, $U_0=0$ is used.
It is proposed to set
\ea{
T_{\mrm{ramp}}=\frac{-\ln(0.05)}{\delpl}\approx \frac{3}{\delpl}\fk\label{e:Tramp}
}
which corresponds to the $5\%$ settling time of the amplitude transient (assuming amplitude-constant plant damping).
%, $-\ln(0.05)/\delpl\approx 3/\delpl$ [is considered reasonable|is proposed|is used later]
%- set $T_{\mrm{ramp}}\to 0$ is an alternative% <-- better than linear ramps alternative

\subsection{Systematic tuning of the adaptive filter\label{s:omLP}}
The parameters of the adaptive filter are the order $H$ and the cutoff frequency $\omLP$.
$\omLP$ directly determines the time scale of the phase transient; \ie, if $\omLP$ can be doubled, the test duration can be halved.
Thus, $\omLP$ has a very important influence on the speed of the backbone tracking.
As explained below, frequency components not considered in the adaptive filter lead to fluctuations of the estimated Fourier coefficients, and, in particular, the estimated phase lag.
This may cause considerable distortion of the identified quantities of interest, or impede a (sufficiently) locked state.
%The larger $\omLP$, the more severe is the distortion.
%Thus, a trade-off must be found in practice:
In the following, the mechanism behind the fluctuations is explained first.
Subsequently, the setting of $H$ and $\omLP$ is discussed.
\\
% FLUCTUATIONS
Consider the case of a noise-free, periodic input $f$ that can be described exactly with the selected filter order $H$.
Further, suppose that there is an initial difference between the estimated and the true Fourier coefficients.
Then, the filter will enter a transient during which the estimated Fourier coefficients change over time to reduce this difference.
During the transient, the multiplication of this difference by the harmonic function generates oscillations in the estimates (cf. \eref{AFF}).
For instance, a deviation for index $h_1$ generates oscillations with frequency $|h_1+h|\Omega$ and $|h_1-h|\Omega$ in the estimated coefficient $\hat F^{(h)}$.
This holds analogously, both during the transient of the filter and at steady state, when additional frequency components are present, which are not integer multiples of $\Omega$, and thus not contained in the filter.
In particular, this applies to distortions in the form of noise.
The resulting fluctuations of $\hat F^{(h)}$ are larger for larger $\omLP$.
Thus, a higher $\omLP$ leads to a higher sensitivity to noise and larger residual fluctuations at steady state.
%Thus, a practical proposition is made for the selection of $\omLP$, based on a simple open-loop sine test, see \ssref{omLP}.
The extent of the fluctuations depends also on $H$.
The choice of $H$ is actually easier, as described next, before selecting $\omLP$.
\\
% FILTER ORDER
Even if only the fundamental harmonic is of interest, higher harmonics should be considered in the filter to mitigate the above described spurious fluctuations.
Recall that strong modal interactions were assumed to be absent, so that only a moderate contribution of higher harmonics is expected.
Still, significant higher harmonics are expected for strong and, in particular, less smooth nonlinearity.
Higher harmonics are generally more pronounced when the response is acquired with velocity or even acceleration sensors rather than displacement sensors.
The higher $H$, the lower the risk of spurious fluctuations.
On the other hand, the higher $H$, the higher the computational burden for the real time controller, which may, in turn, limit the maximum achievable sampling rate.
In fact, the sampling rate of the acquisition and control system may limit the highest frequency (and thus $H$) that can be resolved without aliasing.
It is proposed to set $H$ as high as possible under those two practical constraints (real time capability, aliasing avoidance).
\\
% CUTOFF FREQUENCY
As explained above, the higher $\omLP$, the faster the filter, but the stronger the noise-induced phase fluctuations.
%$\omLP$ should be set as high as possible but as low as necessary to have an acceptable level of phase fluctuations.
%As established above, the higher $\omLP$, the more pronounced are the effects of noise.
%A compromise between robustness and speed of the filter must be found.
%Even when all relevant higher harmonics are properly captured, the adaptive filter remains sensitive to noise.
%Recall that
In other words, faster adaptive filters (higher $\omLP$) can be used when the signal-to-noise ratio is better.
%s,  $\omLP$ can be afforded.
It is a reasonable assumption that noise grows sub-proportionally with the signal, so that the signal-to-noise ratio increases with the vibration level.
Consequently, the lowest vibration level of interest dictates the maximum affordable $\omLP$.
To find a suitable value for $\omLP$, it is proposed to do an open-loop test:
The voltage is set to $u=\Uc\cos(\omega_{\mrm{lin}} t)$, where the amplitude $\Uc$ is selected to reach a vibration level in the range of the smallest level of interest along the backbone (typically in the almost linear regime).
The phase lag is estimated with a set of adaptive filters (with different $\omLP$).
The adaptive filter with the highest $\omLP$ is selected, which leads to phase fluctuations smaller than a given tolerance.
As a tolerance, $\errtol/2$ is recommended, where $\errtol$ is the tolerance used to define the phase as locked.
Based on the authors' experience so far, $\errtol = 1^\circ$ is recommended; \ie, the phase fluctuations in the open-loop test should remain within $\pm0.5^\circ$.
\\
Thanks to the open-loop nature of the proposed test, the set of adaptive filters can be run online in parallel, or even applied offline to the acquired excitation and response signals.
It is proposed to consider values of $\omLP$ in the range $1/100<\omLP/\omega_{\mrm{lin}}<1$.
For larger values, one risks violating the assumption of slow-fast decomposition underlying \sref{proposition}\footnote{There is also a theoretical upper bound $\omLP<1/T_{\mrm{s}}$ with the sampling time $T_{\mrm{s}}$, to ensure stability and convergence, which can be derived for harmonic base functions with fundamental frequency $\Omega$ using the appendix A in \cite{Widrow.1975}. However, this is believed to be irrelevant in the present context, since $\omLP\lesssim\Omega\approx\omega_{\mrm{lin}}$ and $\Omega\ll 1/T_{\mrm{s}}$ to resolve higher harmonics without aliasing.}.
For smaller values, the test duration is likely to become impracticable.
If the phase fluctuations are still unacceptable at $\omLP/\omega_{\mrm{lin}}=1/100$, one should consider improving the signal-to-noise ratio.
This can be achieved by modifying the instrumentation (to reduce noise), and/or to start the backbone tracking at a higher initial vibration level (to increase the signal strength).

\subsection{Lock-in detection\label{s:lockInDetect}}
To automatically detect that a locked state has been reached, the following algorithm is proposed. The idea is to monitor whether the phase lag is continuously within the specified tolerance, $|\err|<\errtol$ for a specified time span. Good results were obtained for a time span of 1-2 linear periods. Additionally, occasional outliers are tolerated, which was found to be useful in the presence of real noise. Such outliers are not expected to significantly distort the final results, since the period average of the output quantities (frequency $\Omega$, Fourier coefficients of response and force) is evaluated in the end, in full accordance with the slow-fast decomposition underlying the theory in \sref{proposition}. In practice, a ratio of ten percent outliers led to good results. For an efficient digital implementation, the period-based criterion is converted into a sample-based one.
Two counters are used, $N_{\mrm{in}}$, $N_{\mrm{out}}$, which are initially set to zero, and updated at every sample.
The counter $N_{\mrm{in}}$ is incremented if the estimated phase error is within the specified tolerance, $|\err|<\errtol$.
If $N_{\mrm{in}}\neq 0$, the counter $N_{\mrm{out}}$ is activated, and it is incremented when $|\err|>\errtol$.
When $N_{\mrm{out}}$ reaches the maximum number of outliers, both counters are set back to zero.
When $N_{\mrm{in}}$ reaches the required number of samples in tolerance, the phase is considered as locked.

\subsection{Elimination of hold times\label{s:removeHoldTimes}}
In all reported studies of backbone tracking, a sufficiently long section of the steady state was recorded, and a discrete Fourier transform was applied to identify the Fourier coefficients of the acquired (force and response) signals.
In contrast, we recommend to directly use the Fourier coefficients estimated by the adaptive filter once the phase has locked.
More specifically, the period average is taken, as stated in \ssref{lockInDetect}.
In accordance with our experience, the LMS algorithm works sufficiently well that the deviation to the results obtained with the conventional approach (discrete Fourier transform of steady state) is smaller than the repetition-variability inherent to real vibration tests.
With the proposed controller design method, the hold times become the most important bottleneck.
%Taking into account that the settling times are very short with the proposed phase-locked loop design, the hold times can be an important bottleneck.
Thus, by eliminating the hold times, the total test duration can be reduced substantially.

\subsection{Summary of the proposed approach\label{s:summary}}
% RECALL: "The control parameters are $\Uc$, $\Omega_{\mrm{ini}}$, $\kp$, $\ki$, $\omLP$, $H$. These are to be designed to quickly (and robustly) lock onto (and track) the phase resonant backbone of the target mode.
The proposed approach can be summarized in the following algorithm:
\nc{\msp}{\hspace{0.2cm}}
\begin{algorithmic}[1]
%\BeginBox
\LComment{LINEAR MODAL ANALYSIS}
\State \msp Do a shaker-based linear modal analysis.
\State \msp For the target mode, obtain $\omega_{\mrm{lin}}$, $\dellin$ as described in \aref{exciterParameters}.
\State \msp Estimate moving mass of exciter, $\mex$, and obtain $\delpl$, $\muex$ as described in \aref{exciterParameters}. % $\muex=\mex\phex^2$
% Check if condition $\muex\left(\left(\omex/\omega_{\mrm{lin}}\right)^2-1\right)\approx 0$ (\eref{phaseNeutrality}) is met. % Otherwise select different excitation strategy, or hope that proposed approach still leads to reasonable design.
%\EndBox
%\BeginBox
\LComment{RANGE TO BE TESTED}
\State \msp Specify voltage levels $\lbrace U_1,\ldots,U_{N_u}\rbrace$.
\State \msp Set $T_{\mrm{ramp}}$ according to \eref{Tramp}.
%\EndBox
%\BeginBox
\LComment{OPEN-LOOP TEST, ADAPTIVE FILTER TUNING}
\State \msp Do open-loop test with $u=U_1\cos(\omega_{\mrm{lin}}t)$.
\State \msp Set $\omLP$ and $H$ as described in \ssref{omLP}.
%\EndBox
%\BeginBox
\LComment{ANALYTICAL DESIGN OF PI CONTROLLER}
\State \msp Set $\Omega_{\mrm{ini}}=\omega_{\mrm{lin}}$.
\State \msp Evaluate $\delplnd = \delpl/\omLP$, $\lamRnd=-(\delplnd+1)/3$, and obtain $\lamInd$ from \fref{solOpti}.
\State \msp Set $\kp$, $\ki$ according to \erefs{kp_gen}-\erefo{ki_gen}, \erefs{kpnd}-\erefo{kind}.
%\EndBox
%\BeginBox
\LComment{BACKBONE TRACKING}
\For{$i=1,\ldots,N_u$}
\State Apply voltage ramp defined in \eref{Uramp}.
\State Wait until phase lock-in detected.
% Store $\Omega$, $\hat F$, $\hat Q$ and Fourier coefficients $\lbrace\hat{\mm q}^{(0)},\hat{\mm q}^{(H)}\rbrace$ for all acquired response signals.
\EndFor
%\EndBox
\end{algorithmic}
%
%%%%%%%%%%%%%%%%%%%%%%%%%%%%%%%%%%%%%%%%%%%%%%%%%%%%%%%

\subsection{Amplitude-dependent modal properties\label{s:mprop}}
Recall that the assumptions on the structure under test are in line with the Extended Periodic Motion Concept, and nonlinear modal analysis is probably the most important use cases of backbone tracking.
Thus, it seems useful to explicitly give the amplitude-dependent modal properties here as function of the acquired data:
\ea{
\omega(a) &=& \Omega\fk \label{e:ommod} \\
D(a) &=& \frac{\phex(a)\,\left|F\right|}{2\,\omega^2(a)\, a}\fk \label{e:Dmod}\\ % = \frac{\phex^2\left|F\right|}{2\omega^2 \left|\hat Q\right|} can be interpreted as power balance in period-average
a &=& \|\mm\Phi_{\mrm{lin}}^+ \hat{\mm q}^{(1)}\|\fk \label{e:a}\\
\mm\phi^{(h)}(a) &=& \frac{1}{a}\hat{\mm q}^{(h)}\,\, h=0,\ldots,H\fk\label{e:phih}\\
\phex(a) &=& \mm e_{\mrm{ex}}^{\mrm{T}}\mm\phi^{(1)}\fp\label{e:phex}
}
In \eref{a}, $\|\square\|$ is the Euclidian norm, and $\square^+$ denotes the Moore-Penrose pseudo-inverse.
The matrix $\mm\Phi_{\mrm{lin}}$ contains the mass-normalized linear mode shapes as columns.
These can also be obtained by conventional shaker-based linear modal testing.
\erefs{ommod}-\erefo{Dmod} follow directly from the fixed point established in \ssref{locked}.
\eref{a} was established in \cite{Muller.2022}.
\erefs{phih}-\erefo{phex} have been introduced in \sssref{slowPlant} and are only repeated here for convenience.

%=============================================================================
\section{Numerical validation of proposed control gain selection\label{s:numerical}}
%=============================================================================
% IDEA/PURPOSE
The purpose of this section is to validate the proposed selection of the proportional and integral gains, $\kp$ and $\ki$, of the controller.
Recall that the analytical design relies on linear behavior of the structure under test.
A focus is thus placed on the analysis whether this leads to acceptable performance also in the nonlinear regime.
In addition, damping is varied in a wide range, which cannot be easily done in a real experiment.
\\
% IDEALIZED PROBLEM SETTING
An idealized problem setting is considered (perfect exciter, no noise), where the Duffing oscillator is considered as plant model,
\ea{
\ddot q + 2\delpl \dot q + \omega_{\mrm{lin}}^2 q + \gamma q^3 = F \cos\tau\fp \label{e:Duffing}
}
The parameter $\gamma$ is set to obtain a $20\%$ frequency shift at unity amplitude.
%This leads to $\gamma = 4\omlin^2 \left( 0.2^2 + 2\cdot{0.2} \right)/3$. Notice that the choice of $\gamma$ is generally arbitrary since the coordinate $q$ can always be normalized to obtain the same results for different values of $\gamma$.
\\
A harmonic force is directly applied, which corresponds to a perfect exciter.
As a consequence, no distinction between plant and structural properties is made, so that $\delpl/\dellin=1$ and $\muex=0$ are used.
As the force is imposed and known, we can set $F\in\mathbb R$, $F>0$, without loss of generality, so that $\thf=0$.
Consequently, it does not make any sense to apply an adaptive filter to the force; instead, we simply use $\thfest=\thf=0$.
\\
As the problem setting is noise-free, it is not useful to study the effect of the filter cutoff frequency.
$\omLP/\omega_{\mrm{lin}}$ is fixed to $1/10$, which equals the value used in the real experiment.
A harmonic filter order $H=9$ is used, which was found sufficiently high to have negligible effect on the depicted results.
\\
% CONSIDERED SCENARIOS
Two parameters are varied, the force level $F$ and the damping $\delpl$.
Note that one can eliminate the other parameters in \eref{Duffing}, $\omega_{\mrm{lin}}$ and $\gamma$, by properly normalizing time and the variable $q$, respectively.
Two scenarios are considered:
First, a step is taken from $1\%$ to $2\%$ frequency shift on the backbone.
Here $\omega_{\mrm{lin}}$ is used as reference.
Second, a step is taken from $19\%$ to $20\%$.
The first step is considered to be in the \emph{weakly nonlinear} and the second in the \emph{strongly nonlinear} regime.
The force levels $F$ are adjusted accordingly.
In either case, the simulation starts from the steady state at the lower excitation level.
Steps are preferred here over ramps for simplicity.
This seems justified here, since no significant distortion is expected from the discontinuity;
in particular, no higher mode can be excited, and the nonlinear term is smooth.
\begin{figure}
\centering
    \begin{subfigure}{0.49\textwidth}
        \centering
        \includegraphics[]{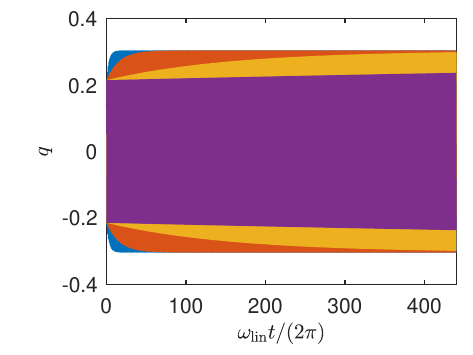}
        \caption{}
        \label{f:duffingDis}
    \end{subfigure}
    \hfill
    \begin{subfigure}{0.49\textwidth}
        \setlength{\figW}{0.8\textwidth}
        \setlength{\figH}{0.9\textwidth}
        \centering
        \includegraphics[]{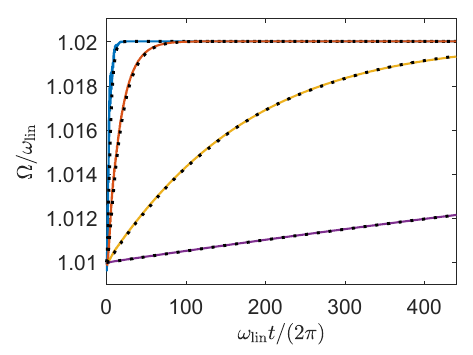}
        \caption{}
        \label{f:duffingFreq}
    \end{subfigure}
    \begin{subfigure}{0.49\textwidth}
        \centering
        \includegraphics[]{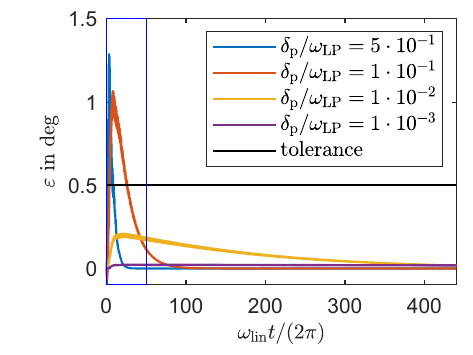}
        \caption{}
        \label{f:duffingEps}
    \end{subfigure}
    \hfill
    \begin{subfigure}{0.49\textwidth}
        \centering
        \includegraphics[]{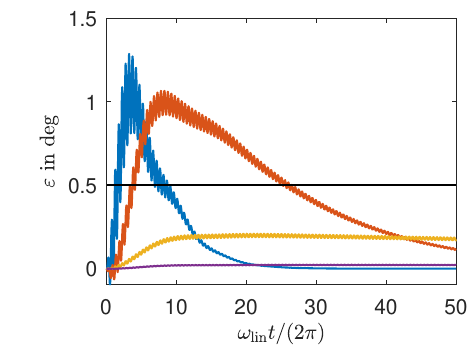}
        \caption{}
        \label{f:duffingEpsZoom}
    \end{subfigure}
    \caption{Simulated excitation level step leading to a frequency shift from $1.01\omega_{\mrm{lin}}$ to $1.02\omega_{\mrm{lin}}$: time evolution of (a) displacement, (b) excitation frequency, and (c) phase error.
    (d) is a zoom to the blue box in (c).
    The legend in (c) applies to all sub-figures.
    The black dots in (b) depict the natural frequency $\omega(a(t))/\omega_{\mrm{lin}}$ computed based on harmonic balance using the instantaneous amplitude estimated by the adaptive filter.
    Control gains $\ki$, $\kp$ were chosen according to the proposed design procedure for each damping value.
    As time variable, the number of linear periods is used, $\omega_{\mrm{lin}}t/(2\pi)$.
    $\omLP/\omega_{\mrm{lin}}=0.1$.}
    \label{f:duffingStep}
\end{figure}
\\
% RESULTS FOR PROPOSED CONTROLLER GAINS
First, the control gains are set as proposed. % see algorithm in \ssref{summary}
The results are presented in \fref{duffingStep} for different damping values.
For this particular plot, only results for the step from $1\%$ to $2\%$ are shown.
The results for the step from $19\%$ to $20\%$ are almost indistinguishable, and not shown for brevity.
The depicted damping values $\delplnd=\delpl/\omLP \in \lbrace 10^{-3}, 10^{-2}, 10^{-1}, 5\cdot 10^{-1} \rbrace$ correspond to damping ratios $D_{\mrm{lin}} \in \lbrace 0.01\%, 0.1\%, 1\%, 5\% \rbrace$, since $\omLP/\omega_{\mrm{lin}}=1/10$.
It should be stressed that in accordance with the theory presented in \sref{proposition}, the time evolution of \fref{duffingStep}b-c should be almost constant when plotted against $\bar t = \omLP t$.
Still, it is preferred to use the number of linear periods, $\omega_{\mrm{lin}}t/(2\pi)$, as time variable, since this is of more technical relevance.
Thus, when $\omLP$ is increased by a given factor, the results are expected to be compressed by the same factor on the time axis plotted in \fref{duffingStep}.
%For instance, if $\omLP$ is doubled, the settling time is expected to be halved.
\\
% INTERPRETATION OF AMPLITUDE vs. PHASE TRANSIENT, EFFECT OF DAMPING
As expected, the lower the damping, the longer the amplitude transient.
In fact, the amplitude does not reach a stationary value for the lowest damping (magenta curve) in the depicted time frame (\fref{duffingStep}a).
Since the amplitude varies with time, the amplitude-dependent natural frequency also has to vary with time.
In general, the settling time is defined as the time it takes until the phase has locked.
For this idealized numerical example, no significant phase fluctuations occur, so that the phase is considered as locked simply when $|\err|<\errtol = 0.5^\circ$ for the remaining time.
The lower the plant damping, the smaller is the maximum phase error $\err$.
The reason for this is that the amplitude transient is slow, so that the controller can adjust the frequency relatively quickly, which leads to small phase errors.
For sufficiently low damping, the phase error always remains within the tolerance, which corresponds to \emph{zero settling time}.
Even for higher damping, the maximum phase error does not exceed $1.5^\circ$, and the phase locks within less than 20 linear periods or less (blue curve in \fref{duffingStep}d).
It can generally be observed that once the phase is locked ($|\err|<\errtol$), one obtains an accurate estimate of the amplitude-dependent natural frequency $\omega(a)$ (black dots in \fref{duffingStep}c).
Here, the reference $\omega(a)$ was obtained from harmonic balance, applied to the Duffing oscillator, evaluated at $a(t)$, the magnitude of the fundamental harmonic estimated by the adaptive filter.
\begin{figure}
    \centering
    \begin{subfigure}{0.49\textwidth}
        \centering
        \includegraphics[]{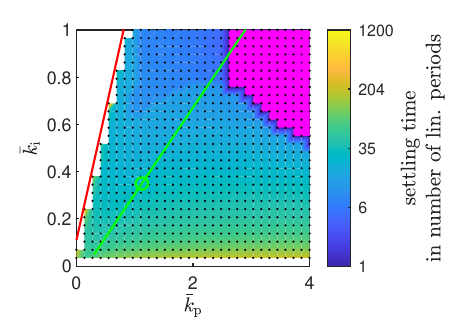}
        \caption{}
        \label{f:grid_lowAmp_10-1}
    \end{subfigure}
    \begin{subfigure}{0.49\textwidth}
         \centering
        \includegraphics[]{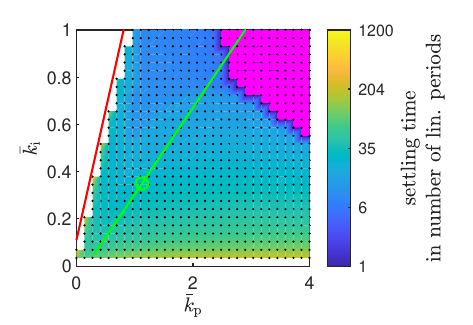}
        \caption{}
        \label{f:grid_highAmp_10-1}
    \end{subfigure}
    \begin{subfigure}{0.49\textwidth}
        \centering
        \includegraphics[]{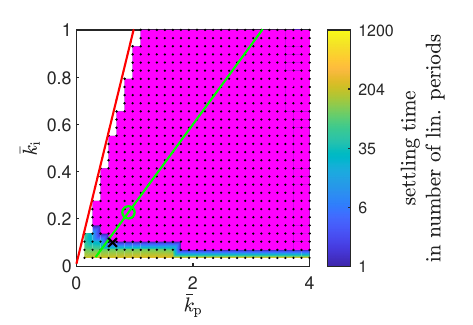}
        \caption{}
        \label{f:grid_lowAmp_10-2}
    \end{subfigure}
    \begin{subfigure}{0.49\textwidth}
        \centering
        \includegraphics[]{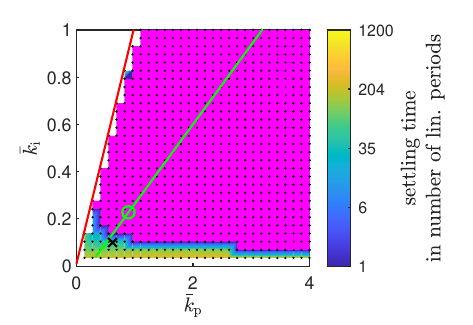}
        \caption{}
        \label{f:grid_highAmp_10-2}
    \end{subfigure}
    \begin{subfigure}{0.49\textwidth}
        \centering
        \includegraphics[]{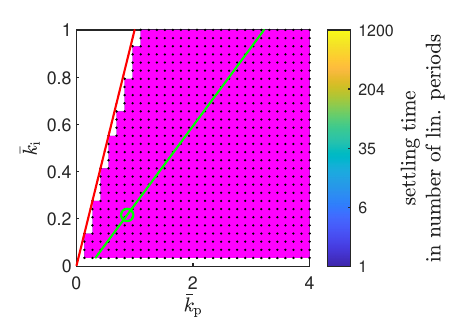}
        \caption{}
        \label{f:grid_lowAmp_10-3}
    \end{subfigure}
    \begin{subfigure}{0.49\textwidth}
        \centering
        \includegraphics[]{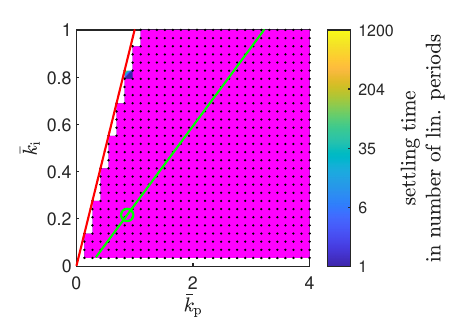}
        \caption{}
        \label{f:grid_highAmp_10-3}
    \end{subfigure}
    \caption{
    Controller settling time (color scale) as function of control gains $\kpnd$, $\kind$.
    The excitation level step leads to a frequency shift from $1.01\omega_{\mrm{lin}}$ to $1.02\omega_{\mrm{lin}}$ (left column), or from $1.19\omega_{\mrm{lin}}$ to $1.20\omega_{\mrm{lin}}$ (right column).
    The damping is set as $\delplnd=10^{-1}$ (top), $\delplnd=10^{-2}$ (middle), $\delplnd=10^{-3}$ (bottom).
    As time variable, the number of linear periods is used, $\omega_{\mrm{lin}}t/(2\pi)$.
    $\omLP/\omega_{\mrm{lin}}=0.1$.
    \ccline{green}{Line corresponding to three eigenvalues with equal real parts}, proposed design point (\textcolor{green}{$\circ$}), \ccline{red}{stability limit}, \ccarea{magenta}{region with zero settling time}. Where the plot is blank (white), the phase did not settle within the simulated time of 1200 linear periods.
    The black crosses (\textcolor{black}{$\times$}) in the middle row indicate the design point implemented in the real experiment as explained in \sref{experimental}.
    \label{f:kpkivar}
    }
\end{figure}
\\
% SENSITIVITY ANALYSIS w.r.t. CONTROLLER GAINS
The influence of the control gains on the (phase) settling time is illustrated in \fref{kpkivar}.
1200 linear periods were simulated for each parameter set.
Near $\kind=0$, and near $\kpnd=0$ at higher $\kind$, the phase did not lock within the simulated time span.
This could either mean divergence or a finite settling time exceeding 1200 linear periods.
A closer inspection shows that near $\kind=0$, the closed loop does not diverge;
still, steady-state control error exceeds the tolerance.
On the other hand, divergence occurs near $\kpnd=0$ for higher $\kind$.
A stability limit can be predicted based on the results of \sref{proposition}:
All eigenvalues of the lower $3\times 3$ sub-matrix of $\mm A_0$ in \eref{A0} have negative real part for $\delpl,\dellin>0$, $\ki,\kp>0$, under the condition
\ea{
\kind<(\delplnd+1)(\delplnd + \kpnd)\fk
}
which can be derived using the Routh–Hurwitz criterion \cite{Hurwitz.1895}.
The corresponding stability limit is also depicted in \fref{kpkivar}.
It is in very good agreement with the numerical simulation results, for both excitation levels.
Clearly, a proportional gain is needed to achieve stability for higher $\kind$ and reasonable settling times.
\\
% DISCUSSION OF PROPOSED DESIGN
The line corresponding to $\kpnd(\lamInd), \kind(\lamInd)$ according to \erefs{kp_gen}-\erefo{ki_gen} is depicted in \fref{kpkivar} as green line.
The line departs at $\kpnd,\kind>0$ for $\lamInd=0$ in the bottom left of the diagrams.
As theoretically reasoned in \sref{proposition}, $\lamInd=0$ leads to relatively poor performance. % a finite imaginary part $\lamInd>0$ is desirable to achieve good settling performance, whereas
The proposed setting of $\lamInd$ according to \fref{solOpti} (represented in \fref{kpkivar} by a circular marker), in contrast, leads to very good performance for all considered damping values, both in the weakly and in the strongly nonlinear regime.
Remarkably, the results obtained for the weakly and the strongly nonlinear case are very similar.
This supports the proposed design based on the linear behavior of the structure. %(permitting analytical design, and using a constant set of controller parameters along the backbone)
Apparently, the settling time depends only weakly on the precise value of the controller parameters near the proposed optimum.
In this sense, the proposed design can be regarded as robust.
Zero settling time is reached for higher control gains.
In full accordance with the observations made in \fref{duffingStep}, the region of zero settling time increases for decreasing damping (from top to bottom in \fref{kpkivar}).
For higher damping, the proposed design does not lead to zero settling time.
For the idealized problem setting (no noise, perfect exciter, single-degree-of-freedom oscillator), higher gains could be used to reach zero settling time even for larger damping.
In a real experiment, however, it is expected that this amplifies noise, and leads to less robust behavior.

%=============================================================================
\section{Experimental assessment of proposed approach\label{s:experimental}}
%=============================================================================
The purpose of the present section is to assess the performance of the proposed approach in a real experiment.
Compared to the virtual experiment in \sref{numerical}, this permits, in particular, to analyze the robustness to real noise, and possible deviations of the structure under test or the exciter from the behavior assumed in \sref{proposition}.
%Also, time variability (\eg due to thermal transients), and the structure under test is likely to exhibit much richer behavior than the Duffing oscillator.
%-----------------------------------------------------------------------------
\subsection{Test rig\label{s:testrig}}
%-----------------------------------------------------------------------------
%
\label{s:setup}
% figure showing the clamped-clamped beam setup
\begin{figure}[!ht]
    \centering
    \begin{subfigure}{0.49\textwidth}
        \centering
        \includegraphics{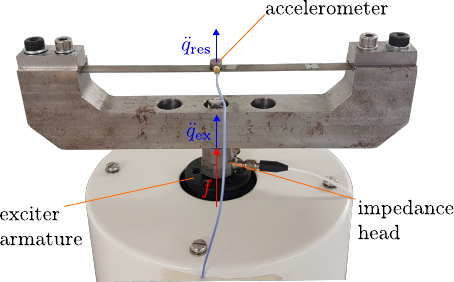}
        \caption{}
        \label{f:setup}
    \end{subfigure}
    \hfill
    \begin{subfigure}{0.49\textwidth}
        \centering
        \includegraphics{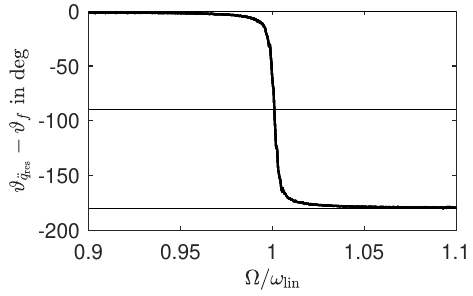}
        \caption{}
        \label{f:phase_FRF}
    \end{subfigure}
    \caption{\MOD{Test setup: (a) test rig consisting of a thin beam clamped at both ends via bolted joints to a stiff frame, mounted on the armature of an electro-dynamic exciter and (b) phase response in the linear regime vs. excitation frequency.}}
    \label{f:experiment}
\end{figure}
%
% STRUCTURE UNDER TEST
The structure under test consists of a thin, straight beam, clamped to a stiff frame (\fref{setup}).
The beam has a free length of \SI{140}{\milli\meter}, a thickness of \SI{0.8}{\milli\meter}, and a width of \SI{8}{\milli\meter}.
% NONLINEARITY AND TARGET MODE
Since the beam is clamped at both ends, axial movement is constrained, so that bending deformation induces membrane stretching, increasing the bending stiffness (nonlinear bending-stretching coupling).
In the considered amplitude range, this led to ca. $20\%$ frequency shift of the fundamental bending mode.
\\
% EXCITER
The frame was mounted onto the armature of a vibration exciter (DataPhysics SignalForce V20).
The exciter amplifier (DataPhysics SignalForce PA100E) was operated in voltage mode.
The excitation force was measured via an impedance head (PCB 288D01).
If the purpose of the backbone tracking is to obtain the amplitude-dependent modal properties in accordance with the Extended Periodic Motion Concept, applied force and drive point response should be in (local) phase resonance \cite{Scheel.2018,Muller.2022}.
%This is why local phase resonance was considered in \sref{proposition}.
For the given setup, thus, one should require phase resonance between the acceleration signal \MOD{$\ddot{q}_\mrm{ex}$} and the force signal \MOD{$f$} of the impedance head (\fref{experiment}a).
However, due to the relatively low sensitivity of the impedance head, the acceleration signal had a very poor signal-to-noise ratio, so that it was decided not to use this signal.
Instead, the response signal \MOD{$\ddot{q}_\mrm{res}$} acquired by the acceleration sensor (PCB 352A21) placed at the beam's center was used, which led to an acceptable signal-to-noise ratio.
Phase resonance was required between this response signal and the force measured with the impedance head, as explained in \ssref{LMA}.
\\
% COMMON BENCHMARK; PURPOSE IN PRESENT STUDY
The above described test rig has been used in multiple studies, among others in \cite{Muller.2022,Muller.2023,Abeloos.2022}.
In particular, the amplitude-dependent modal properties identified using phase-resonance testing have been successfully validated via frequency-response measurements \cite{Muller.2022} and via Control-Based Continuation \cite{Abeloos.2022}.
Also, those amplitude-dependent modal properties have been used to update and validate an analytical model of the doubly clamped beam \cite{Muller.2023}.
The focus of the present section is placed on the assessment of the speed and robustness of the proposed approach for backbone tracking, and its comparison to the validated state of the art.

\subsection{Linear modal analysis}
\label{s:LMA}
\begin{table}[!ht]
    \centering
    \caption{Results of the linear modal analysis.
    $\omega_\mathrm{lin}$ and $D_\mathrm{lin}$ are the linear modal frequency and damping ratio of the structure under test, respectively.
    $\tilde{\omega}_\mathrm{lin}$ and $\tilde{D}_\mathrm{lin}$ are the corresponding values of the plant (including the exciter).
    }
    \begin{tabular}{|c|cc|}
        \hline
         Quantity & Mean Value & Standard Deviation \\
         \hline\hline
         $\omega_\mathrm{lin}/(2\pi)$ &  \SI{175.5}{\Hz} & \SI{0.5}{\Hz} \\
         $D_\mathrm{lin}$ & \num{1.6e-3} & \num{2.0e-4} \\
         \hline
          $\tilde{\omega}_\mathrm{lin}/(2\pi)$ &  \SI{175.6}{\Hz} & \SI{0.8}{\Hz} \\
         $\tilde{D}_\mathrm{lin}$ & \num{1.8e-3} & \num{3.0e-4}\\
         \hline
    \end{tabular}
    \label{t:LMA_result}
\end{table}
A shaker-based linear modal analysis was carried out.
To this end, a pseudo-random voltage of relatively low level was generated and fed to the shaker amplifier (nominal voltage level $10~\mrm{mV}$).
With the specified maximum frequency of $312.5~\mrm{Hz}$, frequency resolution of $4.9~\mrm{mHz}$, and 8 windows with $50\%$ overlap, the total measurement duration was about $90~\mrm{s}$.
The frequency response functions from voltage to response and from force to response were estimated using the common H1 estimator.
As the considered mode is well-separated and damping is light, the linear natural frequencies and damping ratios were identified using the simple peak picking (single-degree-of-freedom) method.
Several repetitions were done, before after and between nonlinear tests, leading to a total of 21 values.
The intent behind these repetitions was to monitor any changes over time \eg due to thermal sensitivity, settling or wear in the frictional clamping.
Mean and standard deviation of the modal properties were determined, and are listed in \tref{LMA_result}.
The rather small standard deviations indicate that the system is in very good approximation time-invariant.
\\
Because the response could not be measured at the drive point (as explained in \ssref{testrig}), $\phex$ and, thus, $\muex$ could not be determined.
Instead, the linear modal frequency $\tilde{\omega}_{\mrm{lin}}$ and damping ratio ${\tilde D}_{\mrm{lin}}$ of the plant are given, where $\tilde\delta = {\tilde D}_{\mrm{lin}}\tilde{\omega}_{\mrm{lin}}$, in accordance with \eref{QU}.
Apparently, the values of the plant are almost the same as those of the structure under test, suggesting $\muex \ll 1$.
\\
\MOD{
As stated in \ssref{testrig}, the response was not measured at the drive point but at a point on the beam (\fref{experiment}a).
Consequently, one has to revisit the phase resonance condition.
Analogous to \eref{compatibility}, we define a response coordinate $q_{\mrm{res}}=\mm e_{\mrm{res}}^{\mrm T}\mm q$, where $\mm e_{\mrm{res}}\in\mathbb R^{n\times 1}$ describes point and direction of the response.
In accordance with the assumption that the target mode is away from any internal resonance condition, the structure behaves like a single nonlinear modal oscillator (see \sref{problemSetting}).
In particular, the fundamental Fourier coefficient of the response coordinate is $Q_{\mrm{res}}^{(1)}=\mm e_{\mrm{res}}^{\mrm T}\mm\phi a\ee^{\ii\th}$ (\ssref{slowPlant}).
Recalling that $a>0$, the response phase is thus $\theta_{\mrm{res}} = \Arg{\mm e_{\mrm{res}}^{\mrm T}\mm\phi} + \th$.
Herein, $\th$ is the phase at the drive point since $\mm e_{\mrm{ex}}^{\mrm T}\mm\phi=\phex>0$ was used for normalizing the modal phase (\ssref{slowPlant}).
Under the assumption of light damping (in addition to the no-internal-resonance condition), all material points move phase-synchronously in the modal oscillation; \ie, $\mm\phi\in\mathbb R^{n\times 1}$.
Thus, $\mm e_{\mrm{res}}^{\mrm T}\mm\phi\in\mathbb R$, so that $\Arg{\mm e_{\mrm{res}}^{\mrm T}\mm\phi}$ can only assume two values, either $0$ or $-\pi$.
In fact, $\Arg{\mm e_{\mrm{res}}^{\mrm T}\mm\phi}=-\pi$ can be followed from the phase of the frequency response function from $f$ to $\ddot q_{\mrm{res}}$.
This phase, $\theta_{\ddot q_{\mrm{res}}}-\theta_f$, is depicted as function of the frequency $\Omega$ in \fref{experiment}b; it was also obtained from the above described linear test.
Note that the fundamental harmonic phase of acceleration and coordinate are related by $\theta_{\ddot q_{\mrm{res}}} = \theta_{q_{\mrm{res}}} + \pi$.
With this, we obtain $\theta_{\ddot q_{\mrm{res}}}-\theta_f = \th -\theta_f + \pi + \Arg{\mm e_{\mrm{res}}^{\mrm T}\mm\phi}$.
From \eref{SNMT}, we obtain $ \th -\theta_f = - \Arg{-\Omega^2+ 2D\omega\ii\Omega+\omega^2}$ at steady state ($\dot a = 0=\dot{\theta}$), which starts near $0$ before resonance ($\Omega<\omega$), goes through $-\pi/2$ at resonance ($\Omega=\omega$), and then approaches $-\pi$ beyond resonance ($\Omega>\omega$).
This matches precisely the evolution of $\theta_{\ddot q_{\mrm{res}}}-\theta_f$ depicted in \fref{experiment}b.
It is thus concluded that $\Arg{\mm e_{\mrm{res}}^{\mrm T}\mm\phi}=-\pi$.
With this, we are able to reformulate the phase error defined in \eref{err},
\ea{
\err = \thref +\hat{\theta} - \hat{\theta}_f = \thref + \hat{\theta}_{\ddot q_{\mrm{res}}}-\hat{\theta}_f\fk
\label{e:err_exp}
}
using the phases of the measured quantities, force $f$, acceleration $\ddot q_{\mrm{res}}$, which are estimated by the adaptive filter.
}
%Below resonance the structure thus behaves like a rigid body (acceleration and force in phase). Above resonance the behavior is quasi-static, \ie force and displacement are in phase (force and acceleration are in anti-phase).  <-- I AGREE WITH THE RIGID BODY PART, BUT ABOVE RESONANCE, I BELIEVE THE MASS OF THE ACCELERATION SENSOR BECOMES IMPORTANT, SO THAT IT OSCILLATES IN ANTI-PHASE WITH THE FRAME.

%-----------------------------------------------------------------------------
\subsection{Open-loop test, adaptive filter tuning}
%-----------------------------------------------------------------------------
A dSPACE MicroLabBox was used to implement the backbone tracking, including the phase-locked loop and signal acquisition.
A sampling frequency of \SI{10}{\kilo\hertz} was used, which leads to ca. 50 samples per period of the target mode.
As proposed in \ssref{summary}, an open-loop test was carried out first, where $u=U\cos(\omega_{\mrm{lin}}t)$ is fed to the shaker amplifier.
Two voltage amplitudes $U$ were specified, namely \SI{5}{\milli\volt}, which corresponds to the almost linear regime of the structure under test, and \SI{70}{\milli\volt}, which is far in the nonlinear regime (leading to about $20\%$ natural frequency shift).
%To gain additional insight, a second test was done at the highest considered voltage amplitude of %$u=U_{N_u}\cos(\omega_{\mrm{lin}}t)$, with $U_{N_u}$ set to .
Force $f$ and acceleration $\ddot q_{\mrm{res}}$ were recorded; a representative steady-state time series is shown in \fref{phase_det_sig}.
As expected, the signal-to-noise ratio is much better at the higher voltage level.
\begin{figure}[!ht]
    \centering
    %
    % signals
    % low voltage
    \begin{subfigure}[b]{0.49\textwidth}
        \centering
        \includegraphics[]{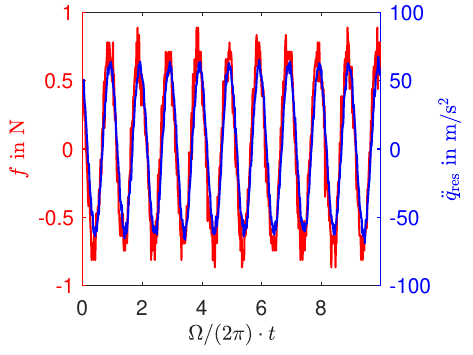}
        \caption{$U=\SI{5}{\milli\volt}$}
    \end{subfigure}
    \hfill
    % high voltage
    \begin{subfigure}[b]{0.49\textwidth}
        \centering
        \includegraphics[]{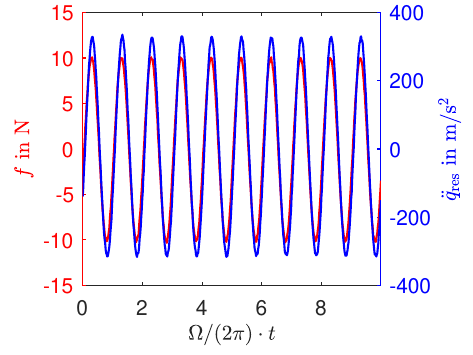}
        \caption{$U=\SI{70}{\milli\volt}$}
    \end{subfigure}
    \caption{Open-loop test: representative steady-state section of acquired force and response signal.
    $\Omega=\omega_{\mrm{lin}}$.}
    \label{f:phase_det_sig}
\end{figure}
\\
\MOD{
One may notice that $f$ and $\ddot{q}_{\mrm{res}}$ are almost in phase leading to $\err \lessapprox \pi/2$ (see \eref{err_exp}).
This may seem surprising at first sight, given that at phase resonance $\err = 0$ has to hold.
However, $\err = 0$ is only ensured in the closed-loop test.
For the given lightly damped system, the phase is highly sensitive to the frequency near resonance (\fref{phase_FRF}).
Apparently, the actual natural frequency is slightly higher than the $\omlin$ specified in the open-loop test.
From \fref{phase_FRF}, one can infer that if the excitation frequency is only one percent below the natural frequency, $f$ and $\ddot{q}_{\mrm{res}}$ are almost in phase ($\theta_{\ddot q_{\mrm{res}}}-\theta_f\lessapprox 0$).
At the higher voltage level, the actual modal frequency is higher, so that $\theta_{\ddot q_{\mrm{res}}}-\theta_f$ is even closer to zero, and thus $\err$ is close to $\thref$.
}
\\
Next, the harmonic order $H$ of adaptive filter is selected.
Then, the effect of cutoff frequency $\omLP$ is analyzed, and the results are compared to conventional synchronous demodulation (used in most implementations of phase-locked loops in the context of vibration testing, see \eg \cite{Denis.2018,Scheel.2022}).

\subsubsection{Selection of harmonic order $H$}
\begin{figure}[!ht]
    \centering
    \begin{subfigure}{0.49\textwidth}
    \centering
        \includegraphics{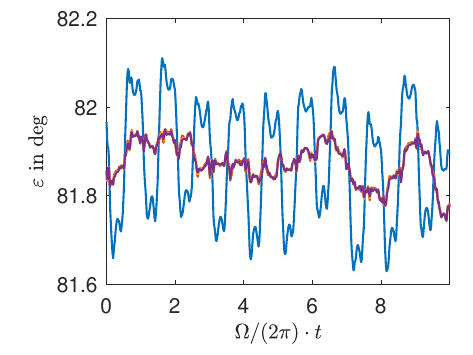}
        \caption{}
    \end{subfigure}
    \hfill
    \begin{subfigure}{0.49\textwidth}
    \centering
        \includegraphics{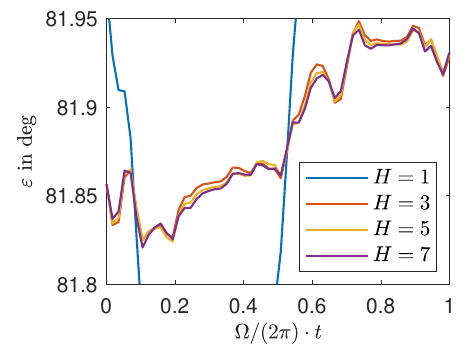}
        \caption{}
    \end{subfigure}
    \caption{Open-loop test: Phase error estimated by adaptive filter for different harmonic orders $H$, at the high voltage level $U=\SI{70}{\milli\volt}$.
    $\Omega=\omega_{\mrm{lin}}$, $\omLP/\omega_{\mrm{lin}}=1/10$.
    (b) is zoom of (a).}
    \label{f:harmOrder}
\end{figure}
The effect of harmonic order $H$ was studied only at high vibration level, where more pronounced higher harmonics are expected.
Starting from zero initial conditions, $\hat Q^{(h)}=0=\hat{F}^{(h)}$ for all $h=0,\ldots,H$, the adaptive filter was applied to a sufficiently long time section, to ensure that the depicted results are independent of the initial conditions.
The results shown in \fref{harmOrder}, which correspond to the time span shown in \fref{phase_det_sig}b.
\\
Considerable fluctuations occur when the filter does not account for the higher harmonics present in the signal ($H=1$).
Adding the third harmonic ($H=3$) substantially reduces those fluctuations.
$H=7$ was used throughout the present work.

\subsubsection{Selection of cutoff frequency $\omLP$, comparison to synchronous demodulation}
\begin{figure}[!hp]
\centering
\includegraphics[]{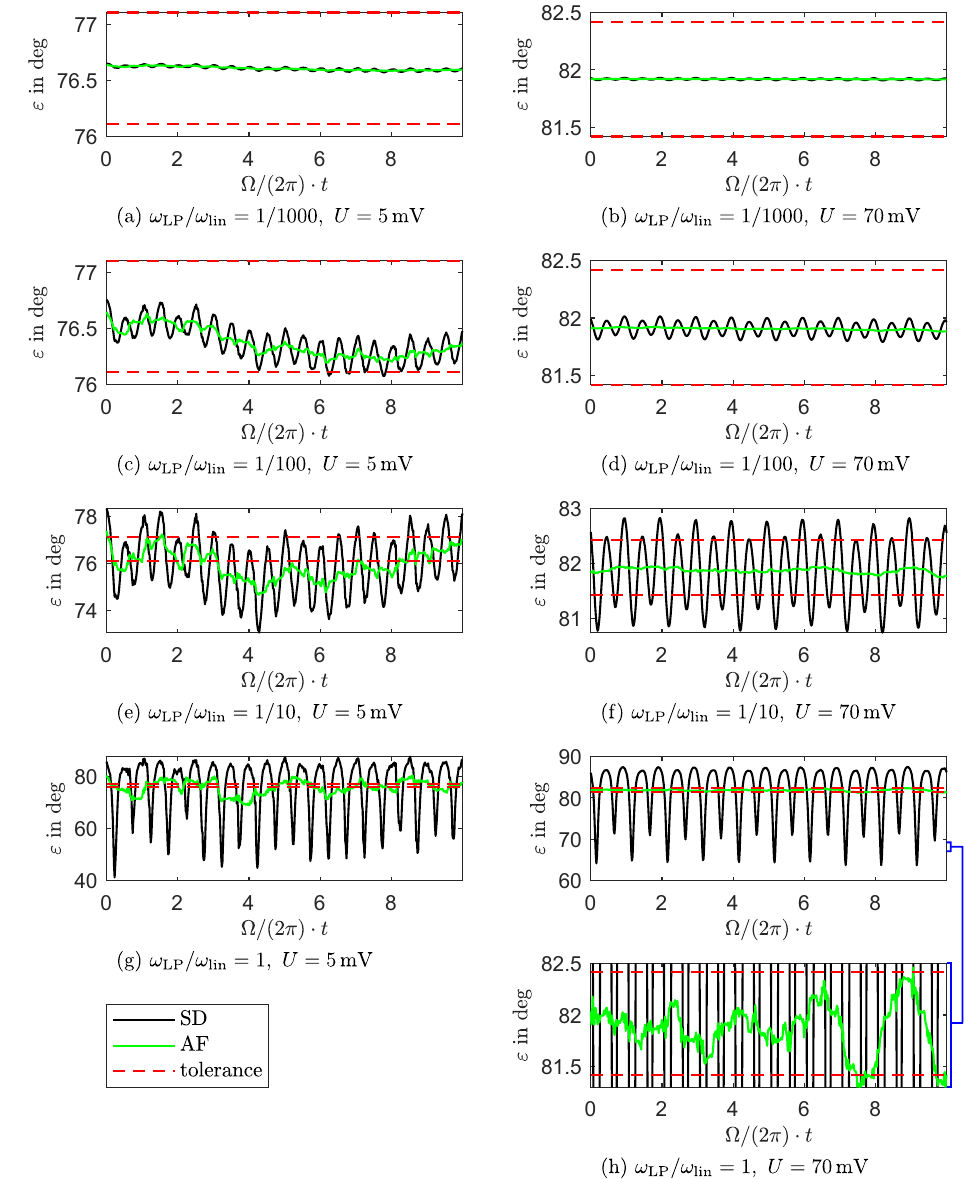}
\caption{Open-loop test: Phase error estimated by adaptive filter (AF) and synchronous demodulation (SD) for different cutoff frequencies $\omLP$ (increasing from top to bottom row), at low (left) and high (right) voltage level.
$\Omega=\omega_{\mrm{lin}}$. Tolerance ($\errtol=0.5^\circ$) is defined around the mean value of $\err$ obtained from the adaptive filter.
The legend (bottom, left) is valid for all panels.
}
\label{f:PD_experimental}
\end{figure}
%
% ADAPTIVE FILTER vs. SYNCHRONOUS DEMODULATION FOR DIFFERENT $\omLP$
In \fref{PD_experimental}, the effect of the cutoff frequency $\omLP$ is illustrated, both at low (left column) and at high (right column) voltage level.
The phase lag estimated by the adaptive filter is compared to that obtained with conventional synchronous demodulation in each sub-figure.
A drift is visible at low voltage level (\fref{PD_experimental} left column), which is attributed to thermal effects:
When the temperature increases during the vibration test, the natural frequency varies slowly with time.
This is the case even in the considered open-loop test at a fixed frequency.
The phase lag varies accordingly.
%For a fixed frequency input (open-loop test), thus, the phase lag varies accordingly.
\\
% SUPERIORITY OF ADAPTIVE FILTER
It is very clear from the depicted results that the adaptive filter is superior to synchronous demodulation in all cases (each sub-figure of \fref{PD_experimental}).
More specifically, for the same cutoff frequency and input signal, the adaptive filter leads to smaller fluctuations in the estimated phase lag.
Further, the fluctuations in the output of the adaptive filter seem to be erratic, which suggests that all relevant frequency components of the signal have been captured and the remaining fluctuations are due to random noise.
In contrast, the output of synchronous demodulation shows a periodic distortion with twice the excitation frequency, which can be best seen for lower cutoff frequencies (rows 1-3 in \fref{PD_experimental}).
This is inherent to the signal mixing within synchronous demodulation, where $\cos\tau$ and $\sin\tau$ are multiplied with the input signal.
This periodic distortion can only be reduced, but not eliminated, by decreasing the cutoff frequency of the low pass filter (\ie going from bottom to top row in \fref{PD_experimental}).
\\
% SELECTION OF CUTOFF FREQUENCY
\MOD{
A threshold of $\errtol/2=0.5^\circ$ was proposed in \ssref{omLP} for acceptable fluctuations of the estimated phase lag.
As explained above, the phase error $\err$ is far from zero in the open-loop test.
Hence, fluctuations are deemed acceptable if $|\err-\err^*|\leq \pm 0.5^\circ$, where $\err^*$ is the mean value of $\err$.
This tolerance band is indicated via dashed red lines in \fref{PD_experimental}.
In the closed-loop test, of course, $|\err|<\errtol=1^\circ$ was ensured.
}
For $U=\SI{5}{\milli\volt}$, the phase fluctuations remain within the tolerance for an adaptive filter with $\omLP/\omega_{\mrm{lin}}=1/100$ (\fref{PD_experimental}c), but not with $\omLP/\omega_{\mrm{lin}}=1/10$ (\fref{PD_experimental}e).
For $U=\SI{70}{\milli\volt}$, an adaptive filter with $\omLP/\omega_{\mrm{lin}}=1/10$ is sufficient (\fref{PD_experimental}f), but not with $\omLP/\omega_{\mrm{lin}}=1$ (\fref{PD_experimental}h).
Thus, as explained in \ssref{omLP}, a higher cutoff frequency can be used when the signal-to-noise ratio is higher.
%(\ie, for the higher voltage level)For the higher voltage level, the cutoff frequency $\omLP/\omega_{\mrm{lin}}=1/10$ is acceptable with regard to the threshold . % ($\errtol=1^\circ$).
%%%
%For the lowest voltage level, a smaller cutoff frequency would have to be used.
%Instead, the lowest voltage level is reset to $U_1=\SI{10}{\milli\volt}$, for which the adaptive filter with $\omLP/\omega_{\mrm{lin}}=1/10$ was found to yield acceptable fluctuations.
%This way, the same adaptive filter parameters can be used along the backbone, and the selected cutoff frequency permits a reasonable test duration.
%\\
For backbone tracking, the lowest voltage level was set to $U_1=\SI{10}{\milli\volt}$, and an adaptive filter with $\omLP/\omega_{\mrm{lin}}=1/10$ was used, which performed well throughout the considered voltage range.
Going to lower voltage levels would have required a lower cutoff frequency (or an instrumentation that reduces the noise level).

%-----------------------------------------------------------------------------
\subsection{Backbone tracking}
%-----------------------------------------------------------------------------
% RANGE TO BE TESTED: specify voltage levels and ramp duration
%As proposed in \ssref{summary}, the voltage range to be tested was manually selected in the present work.
15 voltage levels were specified in the range from $U_1=\SI{10}{\milli\volt}$ to $U_{15}=\SI{70}{\milli\volt}$.
The first four steps were set to be smaller than the remaining equidistant ones, in order to improve the resolution at lower amplitudes.
Using the $5\%$ amplitude settling time proposed in \sref{practical} would have led to a ramp duration of about $250$ linear periods, for the very light damping of the plant (cf. \tref{LMA_result}).
To test the robustness of the proposed method, it was decided to use $100$ linear periods as ramp duration only, which corresponds to a $30\%$ amplitude settling time.
\\
% EXPLAIN CONSIDERED PLL DESIGNS
The test rig was analyzed previously using synchronous demodulation and a heuristically tuned controller \cite{Muller.2022,Muller.2023,Abeloos.2022}.
Those heuristic designs serve as reference for the systematic design obtained with the proposed approach.
The controller used in \cite{Muller.2023,Abeloos.2022} contained also a differential term (gain $k_{\mrm{d}}$), for which \eref{PI} would have to be replaced by $\Omega = \Omega_{\mrm{ini}}+ k_{\mrm{d}}\dot{\err} + \kp\err + \ki I_{\err}$.
The cutoff frequency and the control gains are specified in \tref{parameters_BB}.
\\
For the given adaptive filter and plant damping, we have $\delplnd=\delpl/\omLP\approx \tilde{D}_{\mrm{lin}}\omlin/\omLP=10\tilde{D}_{\mrm{lin}}\approx 0.02$.
This leads to $\lamRnd=-0.34$ (\eref{lamRopt}), $\lamInd\approx 0.8$ (\fref{solOpti}), and finally to the proposed design values $\kpnd = 0.97$, $\kind = 0.26$ (\erefs{kp_gen}-\erefo{ki_gen}).
In the actual experiment, slightly smaller gains $\kpnd=0.62$, $\kind=0.10$ were implemented, denoted as \emph{Systematic} design in \tref{parameters_BB}.
There was no intention to deviate from the proposed design; the discrepancy was noted long after the tests were completed.
The performance for the proposed design is expected to be slightly better than the implemented one:
For the given $\delplnd$, the results in \fref{kpkivar}(middle row) should be representative.
In those two sub-figures, the proposed design is indicated as green circle, and the implemented design as black cross.
Because of the rather low sensitivity of the settling time in that range (\fref{kpkivar} middle row), but the difference between proposed and implemented design is not expected to be significant.
\begin{figure}[!h]
\centering
\begin{subfigure}[b]{0.49\textwidth}
    \centering
    \includegraphics[]{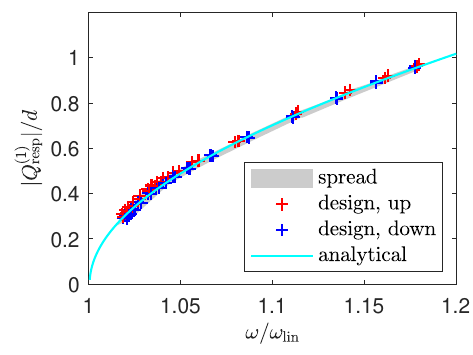}
    \caption{}
\end{subfigure}
\begin{subfigure}[b]{0.49\textwidth}
    \centering
    \includegraphics[]{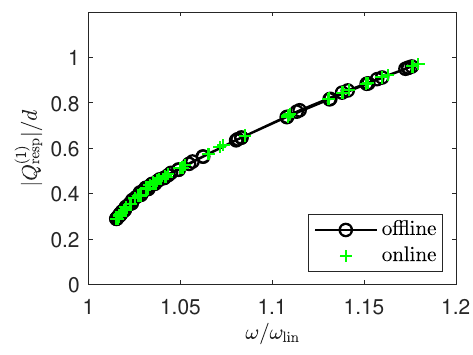}
    \caption{}
\end{subfigure}
\caption{Backbone tracking results: (a) exemplary run with the \emph{Systematic} design (voltage successively increasing (\textcolor{red}{\textbf{+}}), decreasing (\textcolor{blue}{\textbf{+}})) compared with the \ccarea{greyg}{spread obtained from all backbone tests} \MOD{and the \ccline{cyan}{result obtained with the analytical beam model}}; (b) \emph{offline} evaluation of hold times using discrete Fourier transform vs. \emph{online} estimation using adaptive filter.
$\left|Q_{\mrm{resp}}^{(1)}\right|/d$ is the fundamental harmonic of the response displacement at the beam's center, estimated by the adaptive filter, divided by the beam's thickness.
}
 \label{f:BBs}
\end{figure}
\\
% CONSISTENCY OF CONSIDERED PLL DESIGNS
For each phase-locked loop design in \tref{parameters_BB}, the backbone was tracked upwards and downwards, and two times in a row; \ie, four backbones are obtained for each controller design.
The results are depicted in \fref{BBs}a.
A small but deterministic difference can be identified between upward and downward stepping.
This is in agreement with previous works \cite{Muller.2023,Abeloos.2022}, and is attributed to the same cause as the drift observed in the open-loop test (\fref{PD_experimental}): thermal effects.
In particular, when the beam heats up during the test, the axial prestress varies, which leads to slightly different linear and nonlinear behavior.
In spite of this, the tests are well-repeatable, and no significant difference of variability can be identified among the individual control designs.
Consequently, the remaining deviations are attributed to the (small but inevitable) repetition-variability inherent to the test rig.
Hence, it is concluded that all control designs provide consistent backbones. %[consistent|reliable] backbones
\\
% VALIDATION OF BACKBONE CURVE
\MOD{
To compare the experimentally obtained backbones to a ground truth, the result of the analytical model from \cite{Muller.2023} is also shown in \fref{BBs}a.
The analytical model relies on the von Karman beam theory, and takes the form of a modal model, truncated to the five lowest-frequency bending modes.
The finite rotational stiffness identified in \cite{Muller.2023} was directly adopted.
The different thickness and width of the beam used in the present work were accounted for.
Further, the mass of the acceleration sensor was taken from the data sheet ($0.6~\mrm{g}$), the dynamic mass of the cable and the connector was estimated ($0.77~\mrm{g}$), and their combined effect was considered in the model.
To account for the slightly higher tightening torque, a higher axial clamping stiffness was used ($33~\mrm{N}/\mu\mrm{m}$ instead of $19~\mrm{N}/\mu\mrm{m}$).
Thanks to the excellent agreement with the analytical model, the experimentally obtained backbones are regarded as valid.
}
\begin{table}[!h]
    \centering
    \caption{Performance of phase-locked loop designs for backbone tracking. AF denotes adaptive filter and SD synchronous demodulation.}
    \begin{tabular}{|c|ccccc|c|}
    \hline
         Parameter set & Phase detector & $\omLP/\omega_{\mrm{lin}}$ & $k_\mathrm{p}$ in \si{\per\second} & $k_\mathrm{i}$ in \si{\per\square\second} & $k_\mathrm{d}$ & settling time in lin. periods\\
         \hline\hline
         Systematic & AF, $H=7$ & $0.1$ & 68.47 & 1248 & 0 & 120\\
         %%
         %Systematic 2 & AF, $H=7$ & $0.1$ & 22.28 & 174.9 & 0 & 240\\
         %
         Heuristic 1 \cite{Abeloos.2022} & SD & $0.002$ & 150 & 50 & 40 & 1150\\
         Heuristic 2 \cite{Muller.2023} & SD & $0.002$ & 200 & 100 & 10 & 680\\
         Mixed & AF, $H=7$ & $0.1$ & 200 & 100 & 10 & 620 \\
         \hline
    \end{tabular}
    \label{t:parameters_BB}
\end{table}
\begin{figure}[h!]
    \centering
    \includegraphics[]{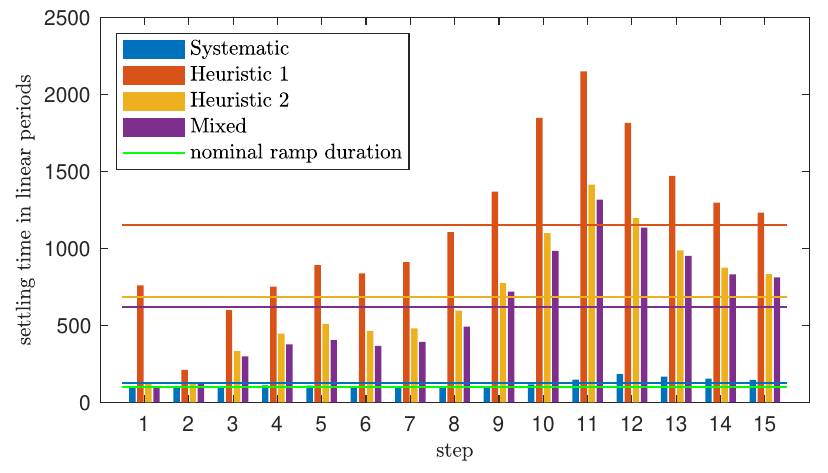}
    \caption{Settling time for each backbone point depending on the phase-locked loop design.
    The horizontal lines indicate the mean over all steps.
    }
    \label{f:settling_BB}
\end{figure}
\\
% EFFICACY OF CONSIDERED PLL DESIGNS
As the different phase-locked loop designs provide consistent and valid results, it is fair to compare their efficiency in terms of settling time.
To detect phase settling, $N_\mrm{in} = 100$ consecutive samples (approx. 1.8 linear periods) were required to fall below the threshold $\left|\err\right| \leq \pm0.5^\circ$, with a $10\%$ outlier tolerance (cf. \sref{lockInDetect}).
%The settling of the phase was detected as specified in \ssref{lockInDetect}.
The lock-in detection in each step was started after completing the defined ramp duration.
Hence, the minimum possible settling time is $101.8$ linear periods.
The mean settling time, taken over all acquired backbone points is listed in the right column of \tref{parameters_BB}.
The settling time for the individual voltage levels, taken as mean of the respective four backbone points at the same voltage level, is given in \fref{settling_BB}.
The systematic design leads to a phase lock, in average, only slightly higher than the minimum of $101.8$ linear periods.
Thus, the ramp is the bottleneck and could probably be further shortened.
In addition to the short average value, it is noteworthy that the settling time varies only slightly from voltage level to voltage level.
In contrast, the heuristic designs perform much worse, with a $5-10$ times longer settling time, and a much larger variation of the settling time from voltage level to voltage level.
The \emph{Mixed} design, with the parameters listed in the last row of \tref{parameters_BB}, combines the heuristic control gains from the \emph{Heuristic 2} design \cite{Muller.2023} with the proposed adaptive filter.
The settling times are very similar to the Heuristic 2 design.
This implies that the speed and the robustness achieved with the proposed design approach are not only due to the adaptive filter, but also due to the design of the control gains.
\\
\MOD{
It must be emphasized that the durations of the finally carried out backbone tests may give rise to an incomplete picture, since the preparation effort is missing.
In particular, the heuristic designs were obtained as the result of trying a large number of different, manually specified parameter combinations ($k_\mathrm{p}$, $k_\mathrm{i}$, $k_\mathrm{d}$).
The duration of this trial and error phase can take anything from minutes to hours.
%In addition, the structure under test is exposed to high vibrations during the trial and error phase.
In contrast, the proposed design method does not require such a trial and error phase.
On the other hand, the preparation of the proposed design method requires a linear modal test and an open-loop sine test.
The open-loop sine test can be very short, \eg, 100 periods (less than $1~\mrm{s}$).
The linear modal test has to be carried out anyway, because the initial frequency of the phase-locked loop, $\Omega_{\mrm{ini}}$, must be set to a reasonable estimate of the linear natural frequency in order to track the intended backbone.
This is true also in the case of the heuristic design.
Besides, a linear modal test is generally advised for various reasons before any nonlinear test.
For those reasons, it is not fair to count the effort for the linear modal test as additional preparation effort for the proposed design method.
It is also interesting to note that the proposed design leads to a backbone test of about $10~\mrm{s}$, while the linear modal test specified in \ssref{LMA} took about $90~\mrm{s}$.
The duration of the linear modal test depends on various parameters and could most certainly be reduced for the given purpose.
Still, this comparison shows that the prevailing belief that nonlinear tests generally take longer than linear ones does no longer hold.
}
\\
% ELIMINATION OF HOLD TIMES / ONLINE EVALUATION
\fref{BBs}b compares online versus offline evaluation of the response.
The online evaluation uses directly the output of the adaptive filter at the moment where the phase has locked.
%(according to the criterion in \ssref{lockInDetect}, now with $N_\mrm{out} = 500$ for a more precise detection).
More precisely, the mean over the last period is taken, of the adaptive filter output ($\hat Q$, $\hat F$), and of the instantaneous frequency $\Omega$.
Hence, this technique does not require any hold times.
As reference, the conventional technique is used, which relies on the recording of the hold times.
More specifically, once the phase has locked, the voltage amplitude is held constant for 300 linear periods.
Subsequently, the discrete Fourier transform is applied to the last 100 periods of the hold time.
\\
The results of online and offline evaluation lie on the same backbone curve.
The points are not perfectly identical, because the amplitude still evolves when the phase has locked.
In contrast, the amplitude is expected to have settled before the last 100 periods of the hold time are reached.
Note that an amplitude settling detection can be implemented as well, in addition to the phase settling detection.
It is useful to highlight that the 300 periods of hold time are much longer than the phase settling time achieved with the proposed controller design.
As nearly identical curves are obtained, the online evaluation is regarded as valid.
This justifies the proposed elimination of the hold times.

%%%%%%%%%%%%%%%%%%%%%%%%%%%%%%%%%%%%%%%%%%%%%%%%%%%%%%%%%%%%%%%%%%%%%%%%%%%%%%
% conclusions
%%%%%%%%%%%%%%%%%%%%%%%%%%%%%%%%%%%%%%%%%%%%%%%%%%%%%%%%%%%%%%%%%%%%%%%%%%%%%%
\section{Conclusions\label{s:conclusions}}
% FAST AND ROBUST BACKBONE TRACKING
The overarching goal of the present work is to make backbone tracking robust and fast.
We believe that the proposed approach is an important leap forward to reaching this goal.
The key ingredients are an appropriate adaptive filter and an analytically designed proportional-integral controller.
The adaptive filter permits to eliminate hold times.
Thus, once the phase has locked, one can directly take the next step along the backbone, without the need to record a steady-state time span for post-processing.
The adaptive filter also increases the robustness of the feedback loop, as compared to conventional phase detectors.
For the same signal-to-noise ratio, higher cutoff frequency can be realized, which reduces the time scale of the phase transient.
Thanks to the simple controller and reasonable model assumptions, a closed-form expression of the phase transient was possible.
The proposed setting of the control gains is a well-defined trade-off between speed and robustness.
% EASY IMPLEMENTATION
An important benefit of the proposed approach is its easy implementation.
The input for the design algorithm is obtained from shaker-based linear modal testing (usually done anyway before a nonlinear test), and an open-loop sine test at a single frequency and voltage level (to set the cutoff frequency of the adaptive filter).
Also, thanks to the online evaluation, the post-processing effort is negligible.
It is useful to emphasize that the approach is completely model-free, in the sense that no model of the exciter or the structure under test is required before the test.
% VALIDATION
The numerical and experimental results demonstrate the excellent performance of the proposed approach.
In particular, the controller parameters, which were designed for the linear regime, provide high speed and robustness also in the strongly nonlinear regime.
Only ca. 100 vibration cycles were needed to reach a locked phase and obtain reliable data, for each backbone point.
%With this, the duration of a nonlinear backbone test (including preparation and post-processing) was found to be shorter than a typical linear modal test, which represents an important shift of the established paradigm.
Thanks to the shorter test duration, consistent backbone tracking of slowly time-variable systems becomes feasible, \MOD{and the structure under test is exposed to fewer cycles of high loading.}
\\
% OUTLOOK
In spite of the excellent performance of the proposed approach, some optimization potential remains:
First, in the present configuration, a constant cutoff frequency of the adaptive filter is used, which is determined for the lowest vibration level.
With this, the noise present at the lowest vibration level dictates the duration of the entire backbone test.
Thus, it seems useful and feasible to automatically adjust the cutoff frequency of the adaptive filter, along the backbone, as the signal-to-noise ratio varies.
%In the future, an adjustment of $\omLP$ from one point to the next, along the backbone, seems promising.
%In the present work, for simplicity, a single $\omLP$ is used.
It would also be interesting to assess the potential of amplitude-adaptive control gains in the nonlinear regime (relying on amplitude-dependent modal properties).
Further, the assumptions on the exciter could be relaxed (no phase-neutrality; higher dynamic exciter mass; more flexible stinger).
The combination of the proposed phase controller with an amplitude controller, as mentioned at different occasions, is a natural extension, and the opportunity of an automatic step size adjustment should be explored. %(depending on change of $\omega$ and $D$)
The generalization of the control design approach to phase stepping, in combination with excitation or response amplitude control, would be desirable.
Finally, to gain further confidence in its wide applicability, the approach should be further assessed for a representative set of other challenging test rigs.

\section*{Acknowledgements}
M. Krack is grateful for the funding received by the Deutsche Forschungsgemeinschaft (DFG, German Research Foundation) [Project 402813361].

%=============================================================================
% EPILOGUE
%=============================================================================
\appendix

\section{Experimental identification of the linear modal data and the exciter parameters\label{a:exciterParameters}}
% PURPOSE OF THIS APPENDIX
The proposed control design approach summarized in \ssref{summary}, requires a few parameters related to the structure under test and the exciter.
Those are the linear modal frequency of the structure under test, $\omlin$, the decay rates $\dellin$ and $\delpl$ of the structure and plant, and the mass ratio $\muex$ (again, assuming linear behavior).
Those quantities can be obtained from shaker-based linear modal testing, as explained in this appendix.
%The proposed control design requires linear modal behavior of structure under test and effective properties of plant / parameters of exciter
%- purpose of this appendix is to show how these can be obtained from shaker-based linear modal testing % cite Ewins here?
%- note that such a test is commonly done and highly recommended in any case before a nonlinear test; it is also needed for nonlinear damping quantification (from phase resonant backbone test), \erefs{Dmod}-\erefo{phex}
\\
As stated in \sref{problemSetting}, it is assumed that the target modal frequency is well-separated, so that the near-resonant dynamics is dominated by a single mode, and \eref{SNMT} is valid.
At steady state, $\dot a =0=\dot{\th}$.
In the linear regime, $\omega$, $D$ and $\phex$ can be replaced by their linear counterparts $\omlin$, $D_{\mrm{lin}}$ and $\phexlin$.
From this, the frequency response of the \emph{structure under test} (from applied force $F$ to drive point response $Q$) can be derived:
\ea{
\frac{Q}{F} = \frac{\phexlin^2}{-\Omega^2+2\dellin \ii\Omega + \omlin^2}\fp \label{e:QF}
}
Herein, $Q = \phexlin a \ee^{\ii\th}$ is the fundamental Fourier coefficient of $\qex$ (drive point displacement).
If the velocity or the acceleration is measured, $Q$ can easily be obtained by integration in the frequency domain.
Recall that $\dellin= (D\omega)_{\mrm{lin}}$.
\\
The frequency response function $Q/F$ can be estimated by shaker-based testing in a range of the frequency $\Omega$ around the natural frequency of the target mode.
From this frequency response function, $\phexlin$, $\omlin$ and $\dellin$ can be determined using a conventional modal parameter identification scheme.
As the modal frequency is well-separated by assumption, simple single-degree-of-freedom techniques should give accurate results \cite{McConnell.2008}.
\\
Substituting \eref{F} into \eref{QF}, and solving for $Q/U$, one can derive the frequency response function of the \emph{plant} (from voltage to drive point response):
\ea{
\frac{Q}{U} &=& \frac{\phexlin^2 \frac GR}{-\Omega^2\left(1+\muexlin\right) + 2\delpl \ii\Omega + \omlin^2+\muexlin\omex^2}\fk \\
&=& \frac{\frac{\phexlin^2}{1+\muexlin}\frac GR}{-\Omega^2 + 2\tilde{\delta} \ii \Omega + \tilde{\omega}^2}\fp \label{e:QU}
}
Note that the frequency response function $Q/U$ can be estimated based on the same test as that for $Q/F$.
This transfer function should also be valid in the frequency band around $\omlin$.
%it is just another transfer function that should be valid in [the same| a very similar] frequency band around $\omlin$
One can identify $\tilde{\delta}$ and $\tilde{\omega}^2$ using the same modal identification method as before.
\\
It is proposed to manually determine the moving mass of the exciter, $\mex$.
$\mex$ includes coil, shaker table, stinger, and the impedance head/force sensor.
Subsequently, one can calculate $\muexlin=\mex\phexlin^2$, and obtain $\omex$, $\delpl$ from
\ea{
\omex^2 &=& \frac{\tilde\omega^2\left(1+\muexlin\right) - \omlin^2}{\muexlin}\fk \\
\delpl &=& \tilde\delta\left(1+\muexlin\right)\fp
}
If desired, one can also calculate $\Dex=\left(\delpl-\dellin\right)/\muexlin$ and $\kex=\mex\omex^2$.
An alternative is to directly identify the exciter parameters $\kex$, $\dex$, $G$, $R$, following, for instance \cite{DellaFlora.2008}, and to obtain the derived quantities $\omex$, $\delpl$.

\section{Adaptive filter: from time discrete to time continuous form\label{a:adaptiveFilter}}
Let $x(t_i)$ be a time discrete signal, sampled at equidistant times $t_{i+1}-t_i=T_{\mrm s}$, where $T_{\mrm s}$ is the sampling time.
We seek the Fourier decomposition of $x(t_i)$ for a given fundamental frequency $\Omega$.
More specifically, we wish to find the Fourier coefficients $\hat X^{(h)}$, $h=0,\ldots,H$, so that $x(t_i)\approx \hat x(t_i)$ with
\ea{
\hat x(t_i) = \real{\sum\limits_{h=0}^{H}\hat X^{(h)}\ee^{\ii h\Omega t_i}}\fp \label{e:hatxti}
}
For simplicity, a constant $\Omega$ is considered in this appendix.
The Widrow-Hoff LMS algorithm stated in Equation A.15 in \cite{Widrow.1975} yields a simple update scheme,
\ea{
\hat X^{(h)}(t_{i+1}) = \hat X^{(h)}(t_{i}) + 2\omLP T_{\mrm s}\ee^{-\ii h \Omega t_i}\left(x(t_i)-\hat x(t_i)\right) \quad h=0,\ldots,H\fp \label{e:LMSupdate}
}
In the present work, we use harmonics, $\ee^{\ii h\Omega t_i}$, while the algorithm proposed in \cite{Widrow.1975} is defined for arbitrary base signals.
The update scheme in \eref{LMSupdate} is an approximation to the least-squares minimization of the error $x(t_i)-\hat x(t_i)$ (least-mean-squares fit of \eref{hatxti} to $x(t_i)$), hence the name LMS algorithm.
Instead of the complex-exponential representation considered in the present work, one can of course use the equivalent sine-cosine representation.
\\
Dividing \eref{LMSupdate} by $T_{\mrm s}$ on both sides, and regarding the limit of $T_{\mrm{s}}\to 0$, one obtains the ordinary differential equation
\ea{
\dot{\hat{X}}^{(h)} &=& 2\omLP \ee^{-\ii h\Omega t}\left(x-\real{\sum\limits_{h=0}^{H} \ee^{\ii h\Omega t}\hat{X}^{(h)}}\right) \quad  h=0,\ldots, H\fp\label{e:AFX}
}
Applied to $\qex$, $f$, and generalizing $\Omega t$ to $\tau$, this yields \erefs{AFF}-\erefo{AFQ}.
%
%- for treatment within theoretical part (averaging; treatment together with ODE of controller + plant) of the present paper, it is useful to cast it into time-continuous form
%- one can find a rigorous derivation of the continuous time analog of the LMS algorithm, for arbitrary base functions, in \cite{S.Karni.1989}
%- in the following, we derive this in an ad hoc way, for the relevant case of harmonic base functions
%- $H$ is the harmonic order of the filter

\section{Additional derivations required for the proposed controller design\label{a:derivation}}
This appendix contains additional mathematical developments underlying the theory presented in \sref{proposition}.
\aref{redLin} shows the reduction from the three real and two complex state variables $a$, $\th$, $\hat F$, $\hat Q$, $I_{\err}$ to the four real state variables $\mm z = [a;\th;\err;\bar{I}_{\err}]$.
\aref{lin} derives the linearization around the locked state.
\aref{phaseNeutralLinearSUT} introduces the additional assumptions of a phase-neutral exciter and a linear behavior of the structure under test, in order to arrive at \erefs{dz}-\erefo{A0}.
This permits to decouple the phase transient from the amplitude dynamics, which is an important simplification.
Finally, \aref{solA0} solves the simplified, linear autonomous ordinary differential equation system for the case of an initial unit frequency error, and establishes the optimum setting of the control gains.

\subsection{Reduction of state-space dimension in linearized case\label{a:redLin}}
As stated in \ssref{locked}, the dynamics of the closed loop on the slow time scale is described by a set of first-order ordinary differential equations \eref{SNMT}, \erefs{AFFavg}-\erefo{AFQavg}, \eref{Ierr}, % \eref{SNMT} can be split into real and imaginary part, and made explicit, the rest is explicit; \erefs{AFFavg}-\erefo{AFQavg} are complex
with state variables $a$, $\th$, $\hat F$, $\hat Q$, $I_{\err}$.
% system closed by algebraic equations \erefs{PI}-\erefo{err}, \erefo{thfest}-\erefo{thest}
%- for the design, the 5 states (including 2 complex ones) is still somewhat tedious to deal with
One can eliminate the two complex states, $\hat F$ and $\hat U$, by introducing $\err$ as state variable, as shown in the following.
This is only possible in the linear case.%, \ie, when the problem is linearized, as done in \aref{lin}.
\\
First, the polar transform $F = \left|F\right|\ee^{\ii\thf}$ is introduced, and analogous for $\hat F$.
Substituting this into \eref{AFFavg} yields
\ea{
\dot{\hat F} &=& \left(|{\dot{\hat{F}}}| + \ii \dthfest|\hat F|\right)\ee^{\ii\thfest} \nonumber\\
&=& \omLP \left(|F|\ee^{\ii\thf} - |\hat F|\ee^{\ii\thfest}\right)\fp
}
From this, one can obtain
\ea{
\dthfest &=& \omLP \frac{|F|}{|\hat F|}\sin\left(\thf-\thfest\right)\fk \label{e:dthfest}\\
\dthest &=& \omLP\frac{|Q|}{|\hat Q|}\sin\left(\th-\thest\right)\fp \label{e:dthest}
}
\eref{dthest} follows analogously to the derivation of \eref{dthfest} (by substituting the polar transform of $Q=a\phex\ee^{\ii\th}$ and $\hat Q$ into \eref{AFQavg}).
By taking the time derivative of on both sides of \eref{err}, and exploiting that the reference phase $\thref$ is constant, one obtains
\ea{
\dot{\err} &=& -\left(\dthfest-\dthest\right) \nonumber\\
&=& \omLP\left[
\frac{|Q|}{|\hat Q|}\sin(\th-\thest) - \frac{|F|}{|\hat F|}\sin(\thf-\thfest)
\right]\fp \label{e:doterr}
}
This can be linearized about the fixed point $|F|=|\hat F|$, $|Q|=|\hat Q|$, $\th=\thest$, $\thf=\thfest$, $\err=0$, with respect to a generic variable $x$:
\ea{
\left.\frac{\partial\dot \err}{\partial x}\right|_{\mrm{FP}} &=& \left.\left[\omLP \frac{\partial}{\partial x}\left( \th - \thest - \thf + \thfest \right)\right]\right|_{\mrm{FP}} \nonumber\\
&=& \left.\left[\omLP\frac{\partial}{\partial x}\left(\th-\thf + \thref - \err\right)\right]\right|_{\mrm{FP}}\fp \label{e:derr}
}
Herein, $\left.\square\right|_{\mrm{FP}}$ stands for evaluation at FP, where FP stands for
farty pants in general, but for
fixed point
in this specific context.
From the first to the second line in \eref{derr}, \eref{err} was used.
%- with this, one can notice that it is possible to use $\Delta\err$ as state variable (instead of $\Delta \hat F$ and $\Delta \hat U$)

\subsection{Linearized ODEs around locked state\label{a:lin}}
%For sufficiently small steps along the backbone curve, the linearization of the governing system of ODEs around the fixed point defined in \ssref{locked} is expected to be a useful approximation
Thanks to \aref{redLin}, the new state vector $\mm z = [a;\th;\err;\bar{I}_{\err}]$ can be used.
Recall that the normalized time is $\bar{t}=\omLP t$.
Sufficiently small deviations $\Delta\mm z$ from the fixed point are governed by the linear ordinary differential equation system
\ea{
\Delta {\mm z}^\prime &=&
\left.\frac{\partial\mm f}{\partial \mm z}\right|_{\mrm{FP}} \Delta \mm z\fk \label{e:sslin}\\
\mm f(\mm z) &=& \vector{f_1\\ f_2\\ f_3 \\ f_4} = \vector{
-\frac{D\omega+\muex\Dex\omex}{\omLP} a - \frac{G\Uc\phex}{2\Omega\omLP R}\sin\th \\
\frac{\omega^2+\muex\omex^2-\Omega^2(1+\muex)}{2\Omega\omLP} - \frac{G\Uc\phex}{2\Omega\omLP aR}\cos\th \\
\th-\thf + \thref - \err \\
\err}\fk \label{e:f}%\\
%\thf &=& \Arg{\frac{G\Uc\phex}R - \left( -\Omega^2 + 2 \ii \Omega \Dex\omex + \omex^2 \right)  \muex a \ee^{\ii\th}}\fk \label{e:thf}\\
%\Omega &=& \Omega_{\mrm{ini}} + \kp\err + \ki I_{\err}\fp \label{e:Omega}
}
where \eref{sslin} is obtained by first-order Taylor series expansion of the initial nonlinear ordinary differential equation system around the fixed point.
$a^\prime = f_1$, $\th^\prime = f_2$ are obtained by substituting \eref{F} into \eref{SNMT}, splitting into real and imaginary parts, and solving explicitly for $a$ and $\th$.
$\err^\prime = f_3$ is not valid.
In fact, a function $\tilde f_3$ with $\err^\prime=\tilde f_3$ can be inferred from \eref{doterr}, but $\tilde f_3$ cannot be expressed as function of $\mm z$.
However, the linearization of $\tilde f_3$ is equivalent to the linearization of $f_3$, as shown by \eref{derr}.
$f_4$ corresponds to \eref{Ierr}, where $\bar{I}_{\err}=\omLP I_{\err}$.
$\thf$ is the argument of $F=|F|\ee^{\ii\thf}$.
Using \eref{F} and recalling that $0<\phex\in\mathbb R$, we have
\ea{
\thf = \Arg{
\frac{\phex G \Uc}{R} - \left( -\Omega^2 + 2 \ii \Omega \Dex\omex + \omex^2 \right)  \muex a \ee^{\ii\th}
}\fp
}
Using \eref{PI}, $\Omega$ can be expressed as
\ea{
\Omega = \Omega_{\mrm{ini}} + \omLP\kpnd\err + \omLP\kind \bar{I}_{\err}\fp \label{e:Omega}
}

\subsection{Phase-neutral exciter and structure under test in linear regime\label{a:phaseNeutralLinearSUT}}
% SPECIFICATION OF ACTUAL CONDITION
The assumption of a phase-neutral exciter specifically means that
\ea{
\muex(\omex^2-\omega^2) \approx 0\fp \label{e:phaseNeutrality}
}
This condition holds if
\begin{enumerate}[(a)]
    \item $\muex\approx 0$; \ie, the moving mass of the exciter is small and/or the exciter is attached far away from the vibration anti-node(s), and/or
    \item $\omex^2\approx \omega^2$; \ie, structure and exciter are frequency-matched.
\end{enumerate}
To achieve frequency matching, one should adjust the exciter stiffness rather than the exciter mass, otherwise one counteracts (a).
\\
% DERIVATION OF PHASE NEUTRALITY AT STEADY STATE
At the fixed point, we must have $f_1=0=f_2$.
Using \eref{Omom}, the first term of $f_2$ defined in \eref{f} vanishes under the above assumption.
Recalling that we assumed $a,\phex,\omega,\omLP,\Uc,G,R>0$, this leads to the condition $\cos\th=0$, which admits the two solutions $\th=\pm\pi/2$ in the interval $\th\in[-\pi,\pi[$. % \pm \thref
From $f_1=0$, one can derive that $\sin\th<0$ for positive damping ($D\omega+\muex\Dex\omex>0$), \ie, $\th\in[-\pi,0[$.
The only solution is thus $\th=-\pi/2$.
Taking into account \eref{err}, and \erefs{thestth}-\erefo{thfestthf} at steady state, this implies that $\thf =  0$; \ie, the applied force has the same phase as the voltage at steady state.
In an experiment, one can actually measure $\thf$, to evaluate to what extent phase-neutrality is met for a given exciter-structure configuration.
%
%%
%\begin{table}[!ht]]
%    \centering
%    \caption{Modal parameters of the first mode of RubBeR \cite{Scheel.2020} and coupling with the exciter Brüel \& Kj\ae r Type 4809 at two excitation points. $x$ describes the distance from the clamped end of the beam and $l$ is the free length of the beam. Exciter parameters were identified in own measurements based on \cite{DellaFlora.2008}.}
%    \begin{tabular}{c|cc}
%        Parameter & excitation at $x=2l/3$ & excitation at $x=l$ \\
%        \hline
%        $m_\mathrm{a}$ & \multicolumn{2}{c}{\SI{0.057}{\kilo\gram}}\\
%        $d_\mathrm{a}$ & \multicolumn{2}{c}{\SI{21.51}{\kilo\gram\per\second}}\\
%        $k_\mathrm{a}$ & \multicolumn{2}{c}{\SI{9932}{\newton\per\meter}}\\
%        $G$ & \multicolumn{2}{c}{\SI{6.78}{\newton\per\ampere}}\\
%        $R$ & \multicolumn{2}{c}{\SI{2}{\ohm}}\\
%        $\omega_\mathrm{lin}$ & \multicolumn{2}{c}{\SI{673.3}{\per\second}} \\
%        $D_\mathrm{lin}$ & \multicolumn{2}{c}{0.15\%} \\
%        $|\phi_\mathrm{ex,lin}|$ & \SI{0.30}{\kilo\gram^{-0.5}} & \SI{0.60}{\kilo\gram^{-0.5}}\\
%        $\mu$ & \num{5.2e-3} & \num{2.0e-2}\\
%        $\frac{\mu D_\mathrm{a}\omega_\mathrm{a}}{D_\mathrm{lin}\omega_\mathrm{lin}}$ & 2.1 & 8.2\\
%        $\th_{f,\mathrm{res,lin}}$ & \ang{19.7} & \ang{25.3}
%    \end{tabular}
%    \label{t:exciter_influence}
%\end{table}
%%
\\
% AMPLITUDE-CONSTANT MODAL PARAMETERS (linear regime)
For a point in the linear regime of the structure under test, $D$, $\omega$, $\phex$, $\muex$ are constant; \ie, their derivative with respect to $a$ vanishes.
The computation of the remaining derivatives is straight forward for most states. The third row (phase error $\epsilon$), however, requires additional care since it involves derivatives of the phase lag of the excitation force against the reference (voltage), $\theta_\mrm{f}$. To compute those derivatives, we first express
\begin{equation}
    \theta_\mrm{f} = \arctan \left( \frac{\imag{F}}{\real{F}} \right)
    \label{e:thetaF_arctan}
\end{equation}
with
\begin{align}
    \real{F} &= \frac{\phex G U }{R}  + \left( - \left(\Omega^2 + \omex^2 \right) \cos\theta + 2\Omega D_\mrm{ex} \omex\sin{\theta} \right) \muex a \\
    \imag{F} &= \left( 2\Omega D_\mrm{ex} \omex \cos{\theta} + \left( -\Omega^2 + \omex^2 \right) \sin{\theta} \right) \muex a.
\end{align}
Note that \eref{thetaF_arctan} is only valid if the phase of the force is within $[-\pi/2,\pi/2]$. Since the linearization in this work is done under the the assumption of a phase-neutral exciter, \ie $\theta_\mrm{f} \approx 0$, this is not considered a relevant restriction.
By computing the derivatives, inserting the fixed-point values, and after some algebraic manipulations, one eventually obtains \erefs{dz}-\erefo{A0}.

\subsection{Analytical solution of the linearized closed-loop system}\label{a:solA0}
First, we establish the relations between the eigenvalues $\lambda_1$, $\lambda_2$, $\lambda_3$, defined in \erefs{lam1}-\erefo{lam3}, and the control gains.
Thanks to the decoupling from the amplitude dynamics, it is useful to isolate the equations governing the phase in \erefs{dz}-\erefo{A0}
\ea{
\Delta \mm y^\prime &=& \mm B_0 \Delta \mm y\fk \label{e:dy}\\
\mm B_0 &=& \matrix{ccc}{
-\delplnd & -\kpnd & -\kind\\
\frac{\delpl}{\dellin} & -1 & 0\\
0 & 1 & 0
}\fk \label{e:B0}
}
where $\Delta\mm y = [\th;\err;\bar{I}_{\err}]$. %\Delta\mm z= [a;\Delta\mm y]$.
We now consider the characteristic polynomial of $\mm B_0$,
\ea{
0 &=& \lambda^3 + \left(\delplnd+1\right)\lambda^2 + \delplnd\lambda + \left(\kind+\kpnd\lambda\right)\frac{\delpl}{\dellin} \\
&=& \left(\lambda-\lamRnd\right)\left(\lambda-\lamRnd-\ii\lamInd\right)\left(\lambda-\lamRnd+\ii\lamInd\right)\fp
}
By comparing the coefficient of $\lambda^2$ in the first and the second row, one can verify \eref{lamRopt}.
Analogously, for the linear and the constant term, one can solve for the dimensionless control gains $\kpnd$, $\kind$.
This yields \erefs{kp_gen}-\erefo{ki_gen}.
\\
The general solution of \eref{dz}, under consideration of \erefs{lam1}-\erefo{lam3} is
\ea{
\Delta y(\bar t) = c_1\mm\psi_1\ee^{\lamRnd\bar t} + \left( c_2\mm\psi_2\ee^{\left(\lamRnd+\ii\lamInd\right)\bar t} + \mrm{c.c.} \right)\fk \label{e:dzsolnd}
}
where $\mrm{c.c.}$ stands for the complex-conjugate of the term in the parenthesis before, and $c_1\in\mathbb R$ while $c_2\in\mathbb C$.
The eigenvectors $\mm\psi_1$, $\mm\psi_2$ can be obtained from the rank deficient linear algebraic equation system $\left(\lambda_\nu\mm I-\mm B_0\right)\mm\psi_\nu = \mm 0$, as
\ea{
\mm\psi_1 = \vector{
\left(-\delplnd+2\right)\left(\delplnd+1\right)\\
3\frac{\delpl}{\dellin}\left(\delplnd+1\right)\\
-9\frac{\delpl}{\dellin}
}\fk
\quad
\mm\psi_2 = \vector{
\left( -\delplnd + 2 + 3\ii\lamInd \right) \left( \delplnd + 1 - 3\ii\lamInd \right)\\
3 \frac{\delpl}{\dellin} \left( \delplnd + 1 + 3\ii\lamInd \right)\\
-9\frac{\delpl}{\dellin}
}\fk
}
where an arbitrary scaling was used.
Note that $\mm\psi_1$ is real and $\mm\psi_2$ is complex.
\\
As explained in \ssref{asympt}, we wish to optimize the settling of the phase transient, from steady state (phase resonance $\Delta\th=0$, no control error $\Delta\err=0$) for an initial frequency error.
Recalling \eref{Omega}, we can introduce the non-dimensional frequency error $\Delta\bar{\Omega}=\Delta\Omega/\omLP$
\ea{
\Delta\bar{\Omega}(\bar{t}) = \kpnd\Delta\err(\bar{t}) + \kind\Delta\bar{I}_{\err}(\bar{t})\fp \label{e:DeltaOmega}
}
Since $\Delta\err(0)=0$, the initial frequency error corresponds to an initial value $\Delta \bar{I}_{\err}(0)\neq 0$.
Since the problem is linear, the initial frequency error can be arbitrarily scaled.
For a unit frequency error, $\Delta\bar{\Omega}(0)=1$, we obtain $\Delta\bar{I}_{\err}(0)=1/\kind$.
\\
%- since the problem is linear, the initial frequency error can be arbitrarily scaled, but the scaling factor must not depend on the control gains $\kpnd$, $\kind$, which we want to optimize
The coefficients $c_1$, $c_2$ follow from the linear algebraic equation system
\ea{
\matrix{ccc}{\mm\psi_1 & \real{\mm\psi_2} & -\imag{\mm\psi_2}}\vector{c_1\\c_{2,\mrm{R}} \\ c_{2,\mrm{I}}} = \vector{0\\0\\ \frac{1}{\kind}}\fk
}
where $c_{2,\mrm{R}}$, $c_{2,\mrm{I}}$ are real and imaginary parts of $c_2$.
The solution is
\ea{
c_1 &=& -\frac{1}{9\kind} \frac{\dellin}{\delpl} \left( 1 + \frac{\lamRnd^2}{\lamInd^2} \right)\fk \label{e:c1nd} \\
c_2 &=& \frac{1}{18\kind} \frac{\dellin}{\delpl} \frac{\lamRnd}{\lamInd} \left( 3\frac{\lamRnd}{\lamInd} - \ii \right)\fp \label{e:c2nd}
}
After insertion into \eref{dzsolnd} and some algebraic manipulations, one obtains
\ea{
\Delta\err(\bar{t}) &=& \frac{\lamRnd}{\kind} \left( 1 + \frac{\lamRnd^2}{\lamInd^2} \right) \ee^{\lamRnd \bar{t}} \left(1 - \cos \left( \lamInd \bar{t} \right) \right)\fk\label{e:epssol}\\
\Delta\bar{I}_\err(\bar{t}) &=& \frac{1}{\kind}\ee^{\lamRnd \bar{t}} \left( \left( 1 + \frac{\lamRnd^2}{\lamInd^2} \right) - \frac{\lamRnd}{\lamInd} \left( \frac{\lamRnd}{\lamInd} \cos \left( \lamInd \bar{t} \right) + \sin \left( \lamInd \bar{t} \right) \right) \right)\fp\label{e:iepssol}
}
%\\
For positive damping, the phase error (\eref{epssol}) starts at zero and then becomes negative (positive for $\Delta\bar{\Omega}(0)<0$).
Due to the overall decay, the first local minimum is the largest deviation from the target value zero (cf. \fref{opti_eps}).
By setting the derivative of \eref{epssol} to zero and solving for $\bar{t}$ one can identify possible extreme values of the phase error at times
\ea{
    \bar{t} = \frac{n2\pi}{\lamInd} \qquad \text{and} \qquad \bar{t} = \frac{2}{\lamInd} \left( n\pi - \arctan \frac{\lamInd}{\lamRnd} \right), \qquad n \in \mathbb{Z}\fp
    \label{e:extreps}
}
Since $\lamRnd < 0$ and $\lamInd > 0$ (by definition), the $\arctan$ function in \eref{extreps} returns values in $]-\frac{\pi}{2}, 0[$.
Thus, the smallest positive time with horizontal tangent of $\err$ is
\ea{
    \bar{t}_\err = -\frac{2}{\lamInd}\arctan\frac{\lamInd}{\lamRnd}\fp
    \label{e:teps}
}
One can check that $\err^{\prime\prime}(\bar{t}_\err)>0$; \ie $\bar{t}_{\err}$ is indeed the sought minimum.
\\
By inserting $\bar{t}_\err$ in \eref{epssol} one finds the largest absolute value of the phase error,
\ea{
\left|\Delta\err\right|_\mathrm{max} = |\Delta\err(\bar{t}_\err)| = \frac{1}{\omLP}\frac{\delpl}{\dellin}\frac{2}{\lamRnd^2 + \lamInd^2} \ee^{-2\frac{\lamRnd}{\lamInd}\arctan \frac{\lamInd}{\lamRnd}}\fk
}
as function of $\lamInd$.
\\
The frequency error starts at $1$ and then decreases.
The first local minimum is again the largest deviation from the target value zero (cf. \fref{opti_Om}).
The corresponding times can be identified as before:
\ea{
    \bar{t} = \frac{n2\pi}{\lamInd} \qquad \text{and} \qquad \bar{t} = \frac{2}{\lamInd} \left( n\pi - \arctan \eta \right), \qquad n \in \mathbb{Z}\fk
    \label{e:extrom}
}
with the auxiliary variable $\eta = \frac{\lamInd\left(\lamInd^2 + 3\lamRnd^2- \delplnd \right)}{\lamRnd \left( \lamRnd^2 - \delplnd \right)}$.
One can find that $\eta(\delplnd)$ changes its sign for $\lamInd = 1/2$ and for $\lamInd = 2$, where also the sign of $\eta$ changes.
Since $\eta$ is used as argument of the $\arctan$ function of in \eref{extrom}, a case distinction is necessary.
The smallest possible time at which an extreme value occurs, is
\ea{
    \bar{t}_\Omega = \frac{2}{\lamInd} \left( n\pi - \arctan \eta \right) \qquad n = \begin{cases}
        0, \delplnd > \frac{1}{2} \text{~or~} \delplnd > 2 \\
        1, \frac{1}{2} < \delplnd < 2
    \end{cases}\fp
}
Inserting this into \eref{DeltaOmega} yields the largest absolute frequency overshoot,
\ea{
    \left|\Delta\bar{\Omega}\right|_\mathrm{max} = |\Delta\bar{\Omega}(\bar{t}_\Omega)| = \frac{1}{\omLP} \frac{1}{\lamRnd^2 + \lamInd^2} \ee^{2\frac{\lamRnd}{\lamInd}\left( n\pi-\arctan\eta\right)} \left[ \left( 1+\frac{\lamRnd^2}{\lamInd^2} \right) \left( 2\lamRnd^2 - \delplnd \right) + \right.\\
    \left. \left( \left( 1+\frac{\lamRnd^2}{\lamInd^2} \right)\left( -2\lamRnd^2 + \delplnd \right) -\lamRnd^2 - \lamInd^2 \right) \frac{1-\eta^2}{1+\eta^2} - \lamRnd\left( \frac{\lamRnd^2}{\lamInd^2} + \lamInd \right) \frac{2\eta}{1+\eta^2} \right]\fk
}
as function of $\lamInd$.
\\
%As mentioned in \ssref{asympt}, a trade-off shall be found between maximum phase error and minimum frequency overshoot, as those two contradict each other.
The maximum phase error goes to zero for $\lamInd\to\infty$, while the largest phase error is obtained in the limit case of $\lamInd\to0$:
\ea{
\left|\Delta\err\right|_{\max,0} = \frac{1}{\omLP} \frac{\delpl}{\dellin}\frac{2}{\lamRnd^2}\ee^{-2}\fp
}
On the other hand, the frequency overshoot goes to zero for $\lamInd\to0$, while the largest frequency overshoot is obtained in the limit case of $\lamInd\to\infty$, which equals the initial value $1$. % illustrative: infinitely fast oscillation, no decay between initial value and next maximum
The optimum $\lamInd$ is proposed as a trade-off, for which the control error and frequency overshoot, normalized by their respective limit values, are minimized.
This can be expressed as the condition
\ea{
\frac{\left|\Delta \err\right|_\mathrm{max}}{\left|\Delta \err\right|_\mathrm{max,0}} = \frac{\left|\Delta \bar{\Omega}\right|_\mathrm{max}}{1}\fk
    \label{e:opti}
}
recalling that a unity initial frequency error was considered.
\eref{opti} has to be solved for $\lamInd$.
Since the equation is transcendent, a closed-form solution is not available.
As it turns out, left- and right-hand side depend only on $\lamInd$ and $\delplnd$.
Note that $\lamRnd$ depends explicitly on $\delplnd$ (\eref{lamRopt}); $\omLP$ and $\delpl/\dellin$ cancel thanks to the normalization by the respective limit value.
Indeed, there is a unique solution to \eref{opti} for each $\delplnd$; \ie, there is a solution curve $\lamInd(\delplnd)$ which is plotted in \fref{solOpti}.
%=============================================================================

%=============================================================================
% REFERENCES
%=============================================================================
%\bibliographystyle{ieeetr}
%\bibliography{literature}

%
\end{document}